\documentclass{aa}  

\usepackage{graphicx}
\usepackage{txfonts}
\usepackage{xcolor}
\usepackage{hyperref}
\usepackage{booktabs}
\usepackage{siunitx}
\usepackage{placeins}

\begin{document}

   \title{A Three-Dimensional Exploration of Magnetic Fields, Rotation, and Shock Revival in a 39\,M$_\odot$ Core-Collapse Supernova Progenitor}
   \titlerunning{Magnetorotational CCSNe in 3D}

   \author{Liubov Kovalenko \inst{1} ,
          Evan O'Connor \inst{1} ,
          Haakon Andresen \inst{1} ,
          Sean M. Couch \inst{2,3,4}
          }
    \authorrunning{Kovalenko et al.}

   \institute{
   \inst{1} The Oskar Klein Centre, Department of Astronomy, Stockholm University, AlbaNova University Center, SE-106 91 Stockholm, Sweden \\
   \inst{2} Department of Physics and Astronomy, Michigan State University, East Lansing, MI 48824, USA\\
   \inst{3} Department of Computational Mathematics, Science, and Engineering, Michigan State University, East Lansing, MI 48824, USA\\
   \inst{4} Facility for Rare Isotope Beams, Michigan State University, East Lansing, MI 48824, USA }

   \date{Received x xx, xxxx; accepted x xx, xxxx}

\abstract
{We present three-dimensional hydrodynamic and magnetohydrodynamic core-collapse supernova simulations of a rapidly rotating, high-compactness $39\,M_\odot$ progenitor to investigate the roles of rotation and magnetic fields in shock revival and outflow morphology. This study is designed to separate neutrino-driven expansion, rotation-induced deformation, and magnetically aided polar outflow within the same progenitor. We evolve three models: a non-rotating hydrodynamic baseline, a rotating hydrodynamic model, and a rotating magnetized model. All three models reach runaway shock expansion within the simulated interval, but with markedly different morphologies and timescales. The magnetized model revives first and develops the clearest bipolar outflow. The rotating non-magnetized model undergoes the latest shock revival and remains comparatively compact at the end of the simulation. The non-rotating model also undergoes shock revival, but subsequently collapses to a black hole about one second after core bounce. In the magnetized model, Maxwell stresses redistribute angular momentum and extract energy from the differential rotation of the protoneutron star, reducing the inner-core spin and helping channel rotational free energy into the emerging polar outflow. Neutrino emission provides an additional, though smaller, angular-momentum sink in both rotating models. We find that rapid rotation and strong magnetic fields can launch an early magnetically aided polar outflow in 3D, while the resulting dynamics remain intrinsically non-axisymmetric. In this extreme progenitor, rotation also provides significant support against prompt black-hole formation, although the longer-term remnant stability remains uncertain beyond the simulated interval.}
   
   \keywords{magnetic fields -- magnetohydrodynamics (MHD) -- stars: massive -- stars: rotation -- supernovae: general}

   \maketitle
%

\section{Introduction}

The end stage of stars with a zero age main sequence (ZAMS) mass above $\sim 8-10\,M_\odot$ is marked by the collapse of the iron core, the formation of a proto-neutron star (PNS), and a complex interplay between neutrinos, magnetic fields, rotation, nuclear physics, hydrodynamics, and general relativity over several seconds, ultimately leading either to a failed supernova or to a successful core-collapse supernova. The associated energies range from canonical $\sim10^{51}\,\mathrm{erg}$ scale of ordinary core-collapse supernova events to $\sim10^{52}\,\mathrm{erg}$ in the most extreme events. Recent progress in simulating these events with state-of-the-art multidimensional computational packages has revealed a complicated landscape of dynamics involving both successful and failed supernovae, and the production of neutron stars, magnetars, and black holes \citep{marek_delayed_2009, lentz_three-dimensional_2015, melson_neutrino-driven_2015, summa_rotation-supported_2018, oconnor_two-dimensional_2018, oconnor_exploring_2018, kuroda_magnetorotational_2020, chan_black_2018, burrows_physical_2024, nakamura_three-dimensional_2025, burrows_channels_2025, eggenberger_andersen_black_2025, janka_long-term_2025}. In many of these progenitor models the dominant mechanism responsible for reviving the stalled shock and driving the subsequent supernova is the neutrino mechanism \citep{bethe_revival_1985}. Convection and turbulence, which develop due to the neutrino heating deep in the postshock region, are critical in aiding shock revival by increasing the dwell time of material in the gain region, expanding the shock radius, and driving turbulent energy dissipation. Recent work has showed that the neutrino mechanism is particularly robust in high compactness progenitors \citep{pan_equation_2018, walk_neutrino_2020, powell_final_2021, burrows_channels_2025, eggenberger_andersen_black_2025, boccioli_neutrino_2025}, and can in some cases achieve sufficiently high heating efficiencies to power shock revival prior to black-hole formation, producing so-called black hole supernovae (BHSNe; \citealt{eggenberger_andersen_black_2025, andersen_black_2026}). By compactness, we refer to the progenitor property defined in \cite{oconnor_black_2011},
\begin{equation}
\xi_M = \frac{M/M_\odot}{R(M_\mathrm{bary}=M)/1000\,\mathrm{km}}\,.
\end{equation}
For a given mass scale, typically taken as $2.5\,M_\odot$ for black hole-forming scenarios, the compactness is high for progenitors that contain that mass within a small radius. High compactness progenitors are typically associated with larger iron cores, higher postbounce mass accretion rates, and shorter black hole formation times (in the absence of shock revival) when compared to lower compactness progenitors.

An alternative mechanism, not relying solely on neutrino heating, is the magnetorotational (MR) channel for driving shock revival and powering the subsequent supernova \citep{leblanc_numerical_1970, bisnovatyi-kogan_explosion_1971}. Unlike the neutrino mechanism, the MR mechanism requires special initial conditions: rapid rotation and sufficient magnetic fields. If this rotation is present, angular momentum conservation during the collapse creates a substantial reservoir of rotational free energy in the newly-formed, differentially-rotating PNS \citep{wheeler_asymmetric_2000, burrows_simulations_2007, dessart_arduous_2012}. Various processes, including linear winding and flux compression, but perhaps also amplification processes like the magnetorotational instability (MRI), are thought to generate a large scale magnetic field early on in the core-collapse process that can be used to convert this rotational energy into magnetic field energy and through magnetic stresses power a dipolar outflow. Such events have the potential to have energies well in excess of typical core-collapse supernovae, perhaps up to $10^{52}$\,erg, and therefore are a candidate mechanism for so-called hypernovae \citep{iwamoto_hypernova_1998}. This mechanism has been demonstrated over the past 20 years with simulations growing in sophistication from the first 2D Newtonian magnetohydrodynamic (MHD) radiation transport models \citep{burrows_simulations_2007}, to recent 3D models \citep{mosta_magnetorotational_2014, halevi_r-process_2018, kuroda_magnetorotational_2020, obergaulinger_magnetorotational_2021, powell_three_2023, shibagaki_three-dimensional_2024, matsumoto_neutrino-driven_2024, shankar_3d_2026} with various approximations for gravity and neutrino transport, various coordinate systems, and varying numerical resolution. For a recent review of magnetohydrodynamic core-collapse supernova simulations, see the chapter by Bernhard M\"uller in \citet{bambi_new_2025}.

The magnetorotational mechanism relies on strong rotation and magnetic fields, but the values needed are at odds with the predictions from stellar evolution theory for typical massive stars. Theory says that magnetic braking in massive stars will slow down rotating cores \citep{heger_presupernova_2005}, a result is consistent with the observed population of pulsar spin rates, but that prevents core angular momentum values that would be needed for the magnetorotational mechanism. However, this is okay because we know that the majority of supernovae are likely not driven via the magnetorotational mechanism. Only a small fraction of observed supernovae have energies in strong excess of $10^{51}$\,erg and therefore the rates suggest that such progenitors must be rare and likely undergo alternative evolutionary paths \citep{heger_presupernova_2005,woosley_progenitor_2006}. For example, chemically homogeneous evolution, where rapid rotation fully mixes the star on the main sequence, can prevent the star from expanding to large radii after the main sequence. This then reduces the efficacy of magnetic braking, prevents angular momentum loss via winds, and can maintain rapid rotation in the core \citep{yoon_evolution_2005, woosley_progenitor_2006}. Naturally, these stars lack hydrogen and often helium, and therefore fall into the category of stripped envelope supernovae. 
Observationally, magnetorotationally driven events in chemical homogeneous stars may present themselves as part of the energy source in Type Ic-BL supernovae. Type Ic-BL supernovae represent hydrogen and helium poor supernovae with broad lines denoting velocities of $\sim (15,000-30,000)\,$km/s and kinetic energies $\sim (7\pm 6) \times 10^{51}$\,erg \citep{taddia_analysis_2019}. Type Ic-BL supernovae are almost exclusively the supernovae associated with long gamma ray bursts (lGRBs) \citep{galama_unusual_1998,kelly_long_2008} and therefore are a further connection between rotation and magnetic fields (thought to be needed in launching a relativistic jet via some mechanism) and energetic core-collapse events. At the same time, recent work has highlighted that some highly energetic SNe Ic-BL may show no clear central-engine signatures, possibly because any relativistic outflow was off-axis or choked \citep{stritzinger_broad-lined_2026}. Observations of Type I super luminous supernovae (SLSN-I) are also candidates for sites of magnetorotational core-collapse. \cite{gomez_type_2024} has shown that typical kinetic energies of SLSN-I are $\sim3\times10^{51}$\,erg, likely too high for neutrino heating alone. Although the leading candidate for the source of the energy in these events is the spin down power of newly formed magnetars, the spin-down conditions (specifically the fast birth spin rate and the strong magnetic field) suggest that the magnetorotationally driven mechanisms may be at play for the initial shock expansion.

An alternative source of the energy powering Type Ic-BL supernovae, and potentially associated lGRBs, are collapsars \citep{woosley_gamma-ray_1993, macfadyen_supernovae_2001}. The term collapsar broadly encompasses scenarios where accretion onto a black hole formed during a core-collapse event leads to the formation of an accretion disk. Accretion onto the black hole may lead to the formation of a jet and subsequent lGRB. Neutrino driven winds, or more likely viscous processes in the disk ultimately driven by magnetic fields, can drive energetic outflows. Recent progress has shown that these outflows arising from an accretion disk have the potential to reach hypernova-like energies, $\mathcal{O}(10^{51}-10^{52})$\,erg/s \citep{just_r-process_2022, dean_collapsar_2024, fujibayashi_supernovalike_2024, crosatomenegazzi_variety_2024}. 

In both scenarios for energetic supernovae discussed above, i.e. in a fully magnetorotationally driven supernova or in a collapsar-like event, the first seconds following the collapse are critical and must be self-consistently simulated as part of a global picture of energetic supernovae and lGRBs. For instance:
\begin{itemize}
    \item[\scriptsize$\bullet$] The stability and survival of the compact PNS (or protomagnetar) depends on the mass accretion history during the first seconds after the collapse, the strength and evolution of the magnetic field, and the angular momentum distribution in the inner core of the progenitor star.  \cite{obergaulinger_magnetorotational_2020, aloy_magnetorotational_2021, obergaulinger_magnetorotational_2022} performed extensive studies in 2D using a candidate magnetorotational progenitors varying the rotation and magnetic field strength and configuration. \cite{obergaulinger_magnetorotational_2021} followed this up with several 3D simulations. In these works, the authors elucidate the paths toward collapsar or protomagnetar formation. Briefly, they find high compactness progenitors, low magnetic field strengths at bounce, and slow rotation all give more favourable conditions for black hole formation, while lower compactness progenitors, high magnetic fields at bounce, and fast rotation favour protomagnetar formation. 
    \item[\scriptsize$\bullet$] In the case of BHSNe or even magnetorotationally driven supernovae leading to collapsars, the early evolution and resulting morphology is critical for the late time evolution as this will dramatically influence the rate of mass accretion into the central region. This will then impact the disk formation time, the accretion rate into the black hole, and the fuel for a potential collapsar-driven supernova or lGRB. 
    \item[\scriptsize$\bullet$] During the first seconds of the collapse is when the nucleosynthesis of the early ejecta in magnetorotationally driven supernovae is determined. Critically the neutrino interactions, heating timescales, and expansion rate of the ejecta off of the PNS sets the electron fraction of the ejecta. Recent work has shown that these environments have the potential to be a site of heavy element nucleosynthesis beyond the iron group \citep{winteler_magnetorotationally_2012, mosta_r-process_2018, halevi_r-process_2018, reichert_magnetorotational_2023, zha_nucleosynthesis_2024, arcones_origin_2023}, although the exact yields are sensitive to the magnetic field strength, configuration, and the specific dynamical evolution.
    \item[\scriptsize$\bullet$] The magnetic field strength and configuration of the remnant is shaped by the first seconds after the collapse. Various magnetic field amplification processes include flux compression during collapse, the magnetorotational instability (MRI) \citep{akiyama_magnetorotational_2003}, and also convective dynamos \citep{thompson_neutron_1993, raynaud_magnetar_2020} are likely at play in developing the magnetic field. At somewhat later times, fallback could generate magnetic field via a Spruit-Taylor dynamo \citep{barrere_new_2022}. The magnetic field is essential in the magnetar-powered mechanism for SLSNe and even in the collapsar case the magnetic field of the protoneutron star before it collapses may be critical for establishing and maintaining the magnetic field needed to produce a lGRB from a collapsar black hole \citep{gottlieb_shes_2024}. 
\end{itemize}

While rotation and magnetic field evolution can be captured under the assumption of axisymmetry, when simulating rapidly rotating and magnetized progenitors capturing the symmetry-free three dimensional dynamics is critical. Rapidly rotating and magnetised progenitors are susceptible to several instabilities that are intrinsically three-dimensional, including rotational instabilities and associated multimessenger signatures \citep{bugli_three-dimensional_2023}, the rotationally induced standing accretion shock instability (SASI) \citep{summa_rotation-supported_2018,walk_standing_2023}, non-axisymmetric low $T/W$ instabilities \citep{takiwaki_insights_2021}, and the kink instability \citep{mosta_magnetorotational_2014}. These can act to enhance the shock expansion in the case of SASI, redistribute angular momentum in the case of low $T/W$ instabilities, or to disruption the outward propagation of the jetted magnetic outflow in the case of the kink instability.

In this paper we simulate, in 3D Cartesian coordinates, a candidate progenitor for an energetic supernova, a $39\,M_\odot$ ZAMS, quasi-chemically homogeneously evolved massive star evolved with MESA from \cite{aguilera-dena_precollapse_2020}. This magnetized and rapidly-rotating progenitor is one of the highest compactness progenitors ever simulated in 3D whether including rotation and magnetic fields or not. Our main goals are to assess the development and stability of the polar magnetic outflow without the constraints of axisymmetry or coordinate singularities, reveal the impact of the magnetic field in driving shock expansion, and assess the stability of the rotating PNS against black hole formation. We accomplish this by performing three 3D simulations of this progenitor. One MHD simulation with magnetic fields and rotation, one HD simulations with only rotation, and one HD simulation without rotation or magnetic fields. The simulations are carried out through collapse and until 600\,ms--1000\,ms after core bounce, when in all simulations a strong shock revival and runaway shock expansion has occurred.

In our magnetic model we see an energetic dipolar outflow driven by the magnetic fields. The magnetic outflows have jet-like features, but due to the kink instability the development of a strong jetted outflow is suppressed. Our rotating, non-magnetic simulation lacks the dipolar outflow of its magnetic counterpart, but develops a neutrino-driven shock revival at later times. In this model, neutrino heating is reduced because rotational support lowers the neutrino luminosities and mean energies. The non-rotating, non-magnetic simulation develops strong neutrino-driven turbulence, unlike the rotating models where convection is suppressed by differential rotation and neutrino heating is weakened by rotation-modified neutrino emission. This gives rise to a strong neutrino-driven shock expansion but after $\sim1\,\mathrm{s}$ of evolution, the PNS, without the added rotational support, collapses into a black hole.

The remainder of this article is as follows. In \S~\ref{sec:methods}, we present our code infrastructure, including the neutrino physics and gravity treatment, characteristics of the simulated progenitor model and the assumed nuclear equation of state. In \S~\ref{sec:results}, we present the complete dynamical evolution of our 3D simulations and discuss key differences between them as a result of the assumed magnetic field and rotation, we assess the outflow formation and stability within our magnetized simulation, and present an overview of the nucleosynthesis as of the end of the simulations. Relevant analysis details and derivations are introduced in the Results section and in the appendices. We summarize and conclude in \S~\ref{sec:conclusions}.

\section{Methods}
\label{sec:methods}

\subsection{Code Infrastructure}
We perform the simulations with a modified version of \textsc{FLASH},4 \citep{fryxell_flash_2000,dubey_extensible_2009}, a publicly available modular multiphysics code with adaptive mesh refinement (AMR). Magnetohydrodynamics is handled by the \textsc{Spark} solver \citep{couch_towards_2021}, a high-order finite-volume method based on fifth-order WENO reconstruction and the HLLC approximate Riemann solver. The magnetic divergence constraint ($\nabla\!\cdot\!\mathbf{B}=0$) is controlled with a cell-centered GLM divergence-cleaning approach that advects and damps magnetic-divergence errors \citep{dedner_hyperbolic_2002, mignone_high-order_2010}. \textsc{Spark} uses columnar "pencil" structures to increase data locality and parallel efficiency.

We evolve the conservative GLM--MHD system \citep{mignone_high-order_2010},
\begin{equation}
  \frac{\partial \mathbf{U}}{\partial t}
  =
  -\sum_{\ell\in\{x,y,z\}}\frac{\partial \mathbf{F}_{\ell}}{\partial \ell}
  +
  \mathbf{S},
  \label{eq:glm_cons}
\end{equation}
with state, fluxes, and source terms
\begin{align}
  \mathbf{U} &=
  \begin{pmatrix}
    \rho \\
    \rho v_d \\
    B_d \\
    E \\
    \psi
  \end{pmatrix},
  \qquad d\in\{x,y,z\},
  \label{eq:glm_U}
  \\[6pt]
  \mathbf{F}_{\ell} &=
  \begin{pmatrix}
    \rho v_{\ell} \\
    \rho v_d v_{\ell} - B_d B_{\ell} + \delta_{d\ell}\left(p+\tfrac{1}{2}B^2\right) \\
    B_d v_{\ell} - B_{\ell} v_d + \delta_{d\ell}\psi \\
    \left(E + p + \tfrac{1}{2}B^2\right) v_{\ell} - (\mathbf{v}\cdot\mathbf{B}) B_{\ell} \\
    c_h^{2} B_{\ell}
  \end{pmatrix},
  \qquad d,\ell\in\{x,y,z\}.
  \label{eq:glm_F}
  \\[6pt]
  \mathbf{S} &=
  \begin{pmatrix}
    0 \\
    \rho g_d \\
    0 \\
    \rho\,\mathbf{v}\cdot\mathbf{g} \\
    -\,c_h^{2}/c_p^{2}\,\psi
  \end{pmatrix},
  \qquad d\in\{x,y,z\}.
  \label{eq:glm_S}
\end{align}
Here $\rho$ is the mass density, $\mathbf{v}$ the velocity, $\mathbf{B}$ the magnetic field, $p$ the gas pressure, and $\mathbf{g}$ the gravitational acceleration. The total energy density is
\begin{equation}
  E = \rho \epsilon + \tfrac{1}{2}\rho v^2 + \tfrac{1}{2}B^2,
  \label{eq:total_energy}
\end{equation}
where $\epsilon$ is the specific internal energy provided by the tabulated nuclear equation of state. 

In Eqs.~\ref{eq:glm_F}--\ref{eq:total_energy}, $\mathbf{B}$ denotes the solver-normalized field used internally by the GLM--MHD system. This field is related to the physical Gaussian-cgs magnetic field by
\begin{equation}
  \mathbf{B}_{\rm code} = \frac{\mathbf{B}_{\rm phys}}{\sqrt{4\pi}} \, .
\end{equation}

Elsewhere in the paper, when discussing magnetic energy densities and forces, we revert to the physical field normalization, for which the magnetic energy density is $e_B = B_{\rm phys}^2/8\pi$ and the Lorentz force carries the usual $1/4\pi$ factor.

The scalar $\psi$ is the GLM cleaning field that advects and damps $\nabla\cdot\mathbf{B}$ errors with hyperbolic speed $c_h$ and parabolic damping set by $c_p$.

We use the \textsc{Spark} solver for \emph{both} magnetized and purely hydrodynamic simulations. In the MHD model we evolve the full GLM–MHD system (Eqs.~\ref{eq:glm_cons}–\ref{eq:glm_S}). In the HD control models we adopt the identical numerics, AMR setup, and microphysics but suppress magnetic effects by setting $\mathbf{B}=0$ and $\psi\equiv0$, which reduces the system to the compressible Euler equations. This allows differences between the MHD and HD models to be attributed to magnetic stresses rather than to changes in the numerical method, grid treatment, or microphysics.

\subsection{Neutrino Physics}

We evolve neutrino radiation field with a multidimensional, energy-dependent, multispecies two-moment transport scheme with an analytic M1 closure, following the same basic FLASH transport framework described by \citet{oconnor_two-dimensional_2018}, here extended to three dimensions. Specifically, we evolve the zeroth and first angular moments of the neutrino distribution for $\nu_e$, $\bar{\nu}_e$, and $\nu_x$. The neutrino moments are evolved in the laboratory frame, while neutrino energies and interaction rates are defined in the fluid frame. The spatial fluxes of the hyperbolic system of moment equations are computed via a Riemann solver, but in the optically thick regime these fluxes are replaced with the diffusion limit \citep{oconnor_open-source_2015}. Neighboring energy groups are explicitly coupled by velocity-gradient terms and gravitational-redshift terms, allowing for energy-space transport alongside the spatial transport. The neutrino emission and absorption source terms are solved implicitly. The resulting exchange of energy, momentum, and lepton number between matter and radiation is included self-consistently during the evolution via operator splitting. 
 
Neutrino--matter opacities are computed with the open-source library \textsc{NuLib}
\citep{oconnor_open-source_2015} and tabulated as functions of $(\rho,\,T,\,Y_e)$ for each neutrino
species and energy group; during the evolution we trilinearly interpolate these tables.
Because the full interaction kernel (including recoil, weak magnetism, many-body, and
medium corrections) is extensive, we do not reproduce the cross sections here; instead
we summarize the included processes and point to \textsc{NuLib} and the references below
for implementation details.

The interaction set follows standard core-collapse supernova microphysics
\citep{bruenn_stellar_1985,burrows_neutrino_2006,oconnor_open-source_2015}
and includes:
\begin{itemize}
  \item[\scriptsize$\bullet$] \textbf{Charged-current absorption and emission on free nucleons}: $e^- + p \rightleftharpoons n + \nu_e$ and $e^+ + n \rightleftharpoons p + \bar{\nu}_e$, including standard recoil and weak-magnetism corrections \citep{horowitz_weak_2002} and final-state nucleon blocking.

  \item[\scriptsize$\bullet$] \textbf{Electron capture on heavy nuclei}: electron capture producing $\nu_e$, treated using the single-nucleus approximation \citep{bruenn_stellar_1985}.

  \item[\scriptsize$\bullet$] \textbf{Isoenergetic scattering}: scattering of all flavours on neutrons, protons, $\alpha$ particles, and heavy nuclei \citep{bruenn_stellar_1985,burrows_neutrino_2006}. For neutral-current scattering on nucleons, we include approximate medium corrections (virial and many-body corrections) that reduce the opacity and can affect the heating conditions \citep{horowitz_neutrino-nucleon_2017,oconnor_core-collapse_2017}.

  \item[\scriptsize$\bullet$] \textbf{Inelastic scattering}: energy-exchanging scattering of all flavours on electrons, including the associated energy redistribution, following the standard treatments \citep{bruenn_stellar_1985}.

  \item[\scriptsize$\bullet$] \textbf{Pair processes}: $e^- e^+ \rightleftharpoons \nu_x \bar{\nu}_x$ and nucleon--nucleon bremsstrahlung for heavy-lepton flavours $\nu_x$, implemented through effective emissivity and absorption rates \citep{bruenn_stellar_1985,burrows_neutrino_2006,oconnor_open-source_2015}.
\end{itemize}

The transport solver and the opacity set serve different functions in the simulation: the former determines the multidimensional propagation and moment evolution of the neutrino radiation field, while the latter supplies the local emission, absorption, and scattering rates used in the source terms.

\subsection{General Relativistic Treatment}
We incorporate general relativistic effects using an effective gravitational potential, case A of \citet{marek_exploring_2006}. While this is not a full general relativistic treatment, it provides a reasonably accurate approximation for capturing the key relativistic corrections relevant to neutron star formation and core-collapse dynamics \citep{muller_new_2013,schneider_equation_2020}.

Because our models include rapid rotation, it is important to note that the case-A potential does not include the rotation-dependent corrections later introduced by \citet{muller_exploring_2008}. Those corrections were designed for very rapidly rotating configurations in which centrifugal forces strongly modify the collapse dynamics. For slow and moderately rotating collapse, however, \citet{muller_exploring_2008} found that the original case-A potential remains close to the general-relativistic result, while the rotation-corrected potential becomes essential mainly in the more extreme rapid-rotation regime. Given that our models do not undergo a centrifugally dominated bounce, we expect this omission to be subdominant for the global collapse and PNS-structure trends discussed here, although it remains a source of systematic uncertainty.

\subsection{Equation of State and Progenitor}

We adopt the 39\,M$_\odot$ progenitor model from the long-gamma-ray-burst and superluminous-supernova progenitor suite of \citet{aguilera-dena_precollapse_2020}. Within that model set, this progenitor is among the most rapidly rotating at collapse, with \(\Omega_c(t_\mathrm{collapse})=0.58\,\mathrm{rad\ s}^{-1}\), and among the most compact, with \(\xi_{2.5}=0.75\). The main sequence evolution is quasi-chemically homogeneous. This evolutionary pathway leads to a hydrogen-free and helium-poor (only $\sim 0.008\,M_\odot$), compact ($R\sim 0.45\,R_\odot$) progenitor at core collapse, with efficient rotational mixing throughout its life. 

For the rotating models, we initialize the angular velocity with the \emph{shellular} profile provided by the \citet{aguilera-dena_precollapse_2020} stellar-evolution model at collapse, mapped to our grid without modification.

In the magnetic–rotational (B1R1) model we prescribe a \emph{poloidal field} by initializing only one component of the vector potential, $A^\phi(r,\theta)$; all other components are set to zero. This guarantees a divergence-free field with $B_\phi(t{=}0)=0$ and $B_{\rm pol}(t{=}0)\!\neq\!0$ (aligned-dipole–like) and aligns the magnetic axis with the rotation axis. Specifically, we adopt a modified aligned dipole
\begin{equation}
A^\phi(r,\theta) \;=\; B_0\,\frac{R_0^3}{R_0^3 + r^3}\, r\sin\theta,
\end{equation}
which yields a nearly constant field strength for $r\ll R_0$ and a dipolar decay $B\propto r^{-3}$ for $r\gg R_0$. With $B_0=5\times10^{11}\,\mathrm{G}$ this configuration produces a field of $\sim 1.8\times10^{12}\,\mathrm{G}$ at the origin. The non-magnetic rotational (B0R1) and non-rotational (B0R0) controls use the identical hydrodynamic progenitor with $\mathbf{B}=0$.

Although the initial toroidal component is absent, a substantial $B_\phi$ builds up \emph{self-consistently} during collapse and after bounce through differential rotation and, where sufficiently resolved, MHD instabilities. Even a purely poloidal seed can be rapidly wound into a toroidally dominated configuration. Starting with $B_\phi{=}0$ avoids imprinting an \emph{ad hoc} toroidal geometry while still allowing the physically expected build-up of $B_\phi$ at early times.

Our use of a relatively large pre-collapse field follows practical and physical considerations common in global core-collapse supernova (CCSN) studies:
\begin{enumerate}
  \item \textbf{Unresolved poloidal amplification.} Global, neutrino-radiation–hydrodynamics simulations cannot yet afford the resolution and duration needed to capture the full process that generates large-scale poloidal flux from small-scale dynamos/MRI. A strong $B_{\rm pol}$ at $t{=}0$ stands in for this rapid amplification so that magnetically aided outflows can emerge within reasonable runtimes.
  \item \textbf{Progenitor field uncertainties.} Pre-collapse amplitudes and, especially, geometries are uncertain in stellar-evolution models. Initializing with $A^\phi$ provides a controlled, divergence-free large-scale topology while leaving $B_\phi$ to arise dynamically from the simulated rotation profile.
  \item \textbf{Exploratory setup.} Because both the seed-field strength and post-bounce magnetic-field amplification are uncertain, B1R1 should be viewed as an exploratory strong-field model rather than a definitive prediction for this progenitor. It tests how the rapid appearance of an organized, large-scale magnetic flux affects shock revival, polar-outflow morphology and stability, and remnant evolution. The B0R1 and B0R0 simulations provide controlled comparisons in which magnetization is removed ($\mathbf{B}=0$) while all other inputs are kept fixed: B0R1 retains rotation, whereas B0R0 removes both rotation and magnetic fields.
\end{enumerate}

\subsection{Numerical Resolution}
\label{subsec:resolution}
We evolve the MHD/HD–neutrino system on a 3D Cartesian domain
$[-2\times10^{9},\,2\times10^{9}]\,\mathrm{cm}$ in each dimension.
Adaptive mesh refinement (AMR) is employed with ten refinement levels. The finest level provides
$\Delta x_{\min}=6.1035\times10^{4}\,\mathrm{cm}\approx0.61\,\mathrm{km}$, which corresponds near the core to an effective angular resolution
$\Delta\theta\equiv \Delta x/r\simeq0.70^{\circ}$ at $r=50\,\mathrm{km}$.

Outside the inner core, we limit refinement once the local angular resolution reaches $\simeq1.5^{\circ}$. In practice, the equatorial plane is resolved with $\Delta x \approx 0.61\,\mathrm{km}$ in the central $\sim 60\,\mathrm{km}$, with a transition to $\Delta x \approx 1.2\,\mathrm{km}$ over the next several tens of kilometers, and to $\Delta x \approx 2.4\,\mathrm{km}$ by $R \sim 100$--$200\,\mathrm{km}$. Because the AMR geometry is Cartesian, these refinement boundaries are only approximately radial. Refinement is triggered by normalized density and pressure gradients; we do not impose explicit shock-following criteria (no forced shock refinement).

We evolve each model from pre-collapse through core bounce and into the early post-bounce phase. The two rotating simulations reach bounce at $\sim314\,\mathrm{ms}$ after the start of the simulations, while the non-rotating simulation reaches bounce slightly earlier, $\sim310\,\mathrm{ms}$ after the onset of the collapse. From bounce, B1R1 evolves for $\sim700\,\mathrm{ms}$ and B0R1 for $\sim600\,\mathrm{ms}$, whereas B0R0 is evolved for about $1\,\mathrm{s}$, at which point it collapses into a black hole. Establishing the ultimate fate of B1R1 and B0R1 would require substantially longer simulation times. The (M)HD system and coupled neutrino transport are evolved on the adaptive simulation timestep, which varies throughout the evolution but is generally of order $1\,\mu\mathrm{s}$; this time resolution is relevant for rapidly varying diagnostics such as the gravitational-wave signal, however, the plot files used for the analysis are written every $1\,\mathrm{ms}$.

We note, however, that the late-time evolution of the B0R1 model shows some anomalous behaviour associated with the neutrino-transport solution. In particular, at late times the model develops an unusually strong separation between the electron-flavour luminosities, with \(L_{\bar{\nu}_e}\) rising well above \(L_{\nu_e}\), together with a corresponding suspicious rise in the net heating rate (see \S~\ref{sec:neutrino}). A similar feature is also present in an auxiliary higher-resolution counterpart of B0R1, which we ran to assess the sensitivity of selected PNS structural diagnostics to resolution, but it is not seen in B1R1 or B0R0. We suspect that this behaviour is related to the particularly large fluid velocities reached in the core of the B0R1 PNS, which exceed \(\sim 10\%\) of the speed of light. While our neutrino transport includes the full velocity dependence of the neutrino-matter interaction terms, the hydrodynamics is not special relativistic. This mismatch may therefore become especially important in the rapidly rotating, non-magnetized model. In the magnetic model, by contrast, the PNS spins down more efficiently due to magnetic stresses and does not reach comparably large rotation-supported velocities. We therefore treat the final $\sim 250\,\mathrm{ms}$ of B0R1 with particular caution and indicate this interval with partial transparency in the figures. Nevertheless, we include these results for completeness and qualitative comparison, while avoiding overly strong conclusions based on this late-time phase.

\section{Results}
\label{sec:results}

\begin{figure*}[t]
\begin{center}
\includegraphics[width=\textwidth]
{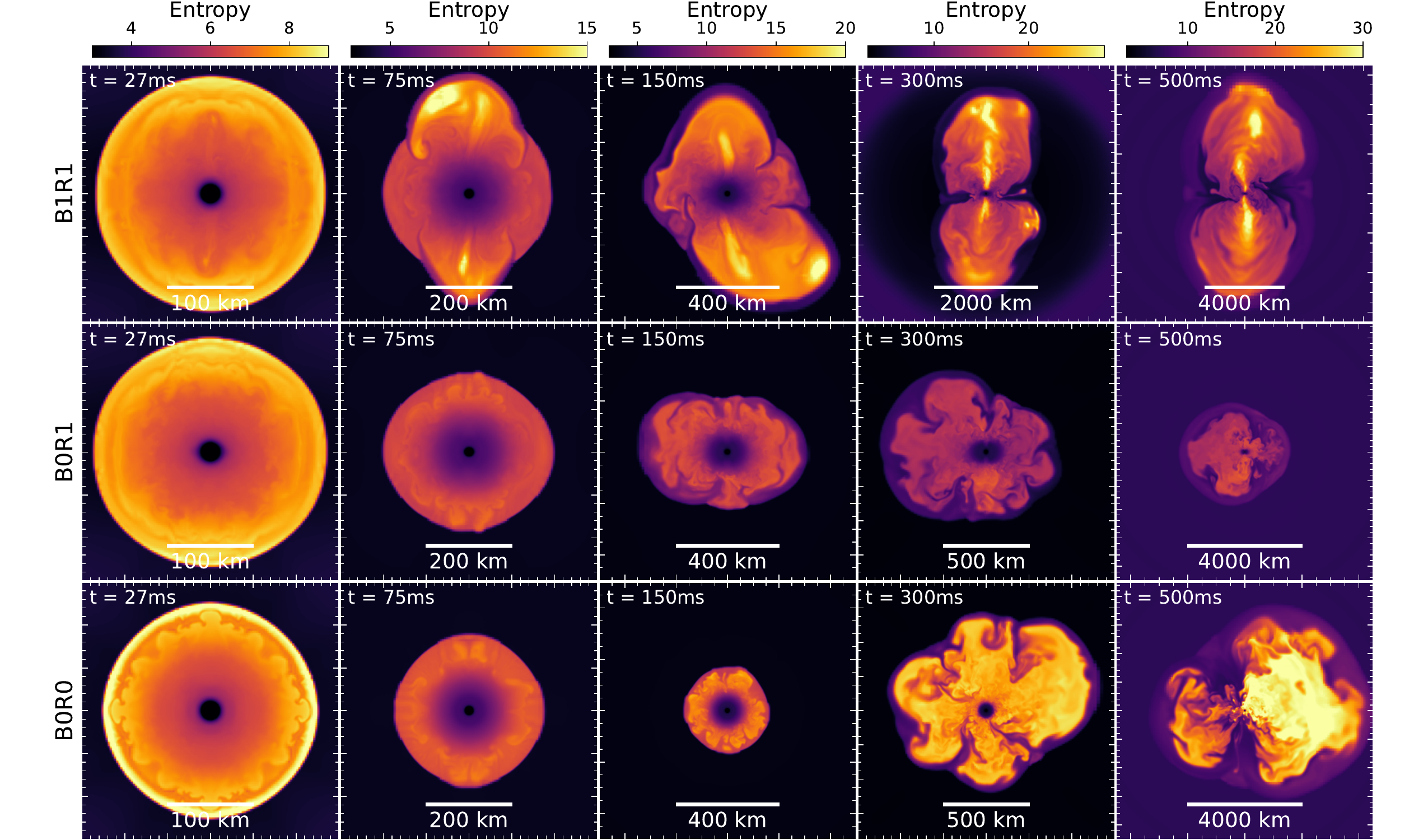}\\
\end{center}
\caption{Entropy slice plots through our three simulations (the y-x plane; top row: B1R1, middle row: B0R1, bottom row: B0R0) for five different post bounce times (from left to right, 27\,ms, 75\,ms, 150\,ms, 300\,ms, and 500\,ms). For B1R1 and B0R1, the rotation axis (+y) is up. The entropy colour scale is fixed across each column for comparison purposes. The spatial scale is also fixed across each column, except for B1R1 at 300\,ms and 500\,ms for clarity.}
\label{fig:slices}
\end{figure*}

\begin{table*}[t]
\centering
\small
\caption{Summary of simulated models and their characteristic times, radii, and diagnostic quantities. $\Omega_{\rm c}$ and $B_{\rm c}$ denote the \emph{central} angular velocity (in $\mathrm{rad\ s^{-1}}$) and \emph{central} magnetic-field strength (in G) of the pre-collapse progenitor used to initialize each model. Here $t_{\rm b}$ is the time of core bounce measured from the start of the simulation, while $t_{\rm end}-t_{\rm b}$ and $t_{\rm rev}-t_{\rm b}$ denote, respectively, the end time of the simulation and the onset time of sustained shock revival measured relative to core bounce. The column "BH" indicates whether a black hole forms \emph{by the end of the simulation}; models marked "no" may still undergo collapse at later times if evolved longer. $M_{\rm PNS,b}$ is the baryonic PNS mass, where the PNS is defined as matter with $\rho \ge 10^{11}\,\mathrm{g\ cm^{-3}}$; the sensitivity of this definition to alternative density thresholds is discussed in \S~\ref{subsec:pns_mass}. The listed PNS masses and radii, the shock radii, and diagnostic energies are measured at $t_{\rm end}$ for each model. Reported polar radii are averages over the north and south poles, and equatorial radii are averages taken around the equatorial plane. For B0R1, the late-time evolution after the onset of the neutrino-transport issue should be interpreted with caution; we nevertheless list the final values for completeness.}

\label{tab:model_summary}
\begin{tabular}{lcccccccccccc}
\toprule
Model &
$\Omega_{\rm c}$ &
$B_{\rm c}$ &
$t_{\rm b}$ &
$t_{\rm end}-t_b$ &
$t_{\rm rev}-t_b$ &
BH &
$M_{\rm PNS,b}$ &
$R_{\rm PNS}^{\rm pole}$ &
$R_{\rm PNS}^{\rm eq}$ &
$R_{\rm sh}^{\rm pole}$ &
$R_{\rm sh}^{\rm eq}$ &
$E_{\rm diag}$ \\
&[$\mathrm{rad\ s^{-1}}$] &[G] & [ms] & [ms] & [ms] & & [$M_\odot$] & [km] & [km] & [km] & [km] & [$10^{51}$ erg] \\
\midrule
B0R0 & 0 & 0 & 310 & 985 & 250 & yes & 2.52 & 24 & 24 & 9003 & 9732 & 2.04\\
B0R1 & 0.58 & 0 & 314 & 571 & 370 & no & 2.38 & 54 & 85 & 2758 & 2933 & 1.44\\
B1R1 & 0.58 & $1.8\times10^{12}$ & 314 & 672 & 150 & no & 2.17 & 43 & 65 & 9440 & 5472 & 3.11\\
\bottomrule
\end{tabular}
\end{table*}

In this section we compare three simulations of the same $39\,M_\odot$ progenitor: the magnetized rotating model B1R1, the rotating non-magnetized model B0R1, and the non-rotating non-magnetized model B0R0. A summary of the models and key times and length scales is provided in Table~\ref{tab:model_summary}. To aid the presentation of the results, we show entropy slices in Fig.~\ref{fig:slices} for each model (B1R1, B0R1, and B0R0 in the top, middle, and bottom rows, respectively) at five post-bounce times: 27, 75, 150, 300, and 500\,ms. For ease of comparison, in each column we keep the colour range on the entropy fixed, although it varies between the columns.

We first present the global evolution of the shock and outflow morphology (\S~\ref{sec:shock}), including the onset time of sustained shock expansion, the degree of asphericity, and the time evolution of diagnostic outflow energies. We then provide baseline neutrino diagnostics relevant for shock revival (\S~\ref{sec:neutrino}), focusing on the luminosities and the net gain-region heating rate. Next, we describe the evolution of the PNS (\S~\ref{sec:pns}), including its mass growth and contraction/oblateness. We then quantify angular-momentum transport and spin evolution in the rotating models (\S~\ref{sec:angmom}), which provides a bridge to the magnetic-field evolution. Finally, we analyze the formation and 3D stability of the magnetically driven polar outflow in B1R1 (\S~\ref{sec:jets}), including kink signatures, tilt, and magnetic-energy budgets, and we summarize broad ejecta-composition properties using post-processed ejecta electron-fraction distributions (\S~\ref{sec:nuc}). 

\subsection{Global shock evolution and morphology}
\label{sec:shock}

All three models undergo a similar collapse and bounce, with a prompt shock launched as the central density exceeds nuclear saturation ($\rho \sim 2.7\times10^{14}\ \mathrm{g\ cm^{-3}}$). At the earliest shown time ($t \approx 27\,\mathrm{ms}$; Fig.~\ref{fig:slices}, left column), the post-shock structure remains fairly similar in all three models and is still close to spherical, although B1R1 already shows the first hint of a jet-like polar feature. Even at this stage, however, prompt convection is more strongly suppressed in the rotating models, especially in the equatorial direction, where differential rotation inhibits the growth of convective motions \citep{fryer_corecollapse_2000}. By $t\approx 75$\,ms, differences are becoming more pronounced. In B1R1, the entropy distribution already now shows a clear north--south elongation, whereas B0R1 and B0R0 remain nearly spherical.

\begin{figure*}[t]
    \centering
    \includegraphics[width=\linewidth]{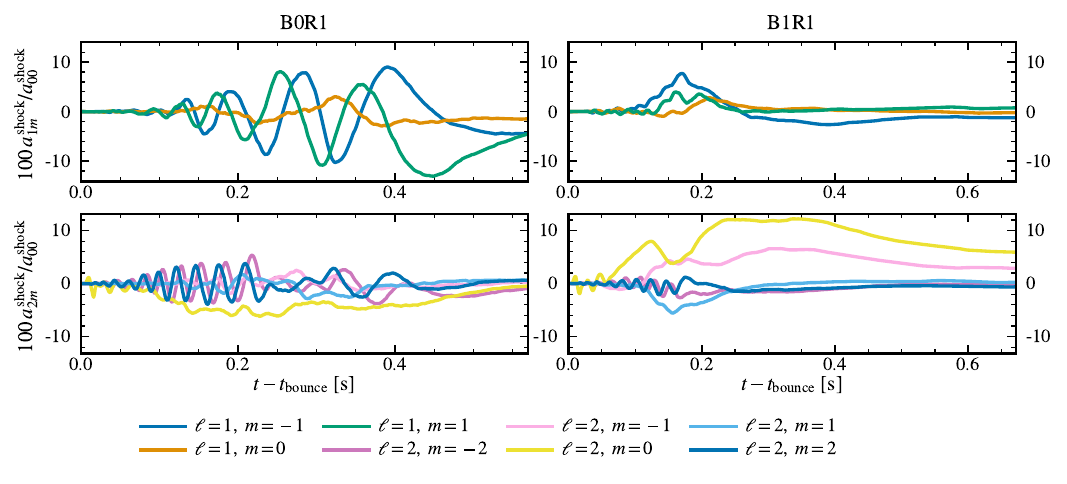}
    \caption{Evolution of the normalized spherical-harmonic shock-deformation amplitudes, $100\,a_\ell^m/a_0^0$, as a function of time after bounce for models B0R1 (left) and B1R1 (right). We show the dominant low-order modes $(\ell,m)=(1,0)$, $(1,\pm1)$, $(2,2)$, and $(2,0)$. The top panels show $\ell = 1$ modes and the bottom panels show $\ell = 2$. In B0R1, the SASI first appears predominantly in $\ell=2$ modes and later transitions to stronger $\ell=1$ activity. In B1R1, the early transient deformation is followed by a short-lived $\ell=2$ SASI episode, after which the magnetically driven bipolar outflow establishes a large-scale, slowly varying asymmetry that suppresses sustained oscillatory behaviour.}
    \label{fig:sasi_modes}
\end{figure*}

By $t\approx 150$\,ms, the divergence between the models is even more apparent. B1R1 has developed a distinctly elongated bipolar morphology, while B0R1 remains only mildly aspherical. B0R0 is still relatively compact at this stage, without the pronounced polar extension seen in B1R1, or the equatorial broadening of B0R1. To quantify the large-scale shock geometry beyond what is visible in the entropy slices alone, we decompose the shock surface into spherical harmonics following the standard CCSN approach \citep{couch_high-resolution_2014, foglizzo_instability_2007}. This decomposition, shown in Fig.~\ref{fig:sasi_modes} shows that the rotating models, especially B0R1, develop strong low-order non-axisymmetric structure. The clearest signatures appear in B0R1: an early quadrupolar $\ell=2$ spiral mode is excited and then gives way to a strong dipolar $\ell=1$ mode as the shock radius grows, as expected from the SASI scaling with shock size \citep{foglizzo_instability_2007}. The $m=\pm1$ components are especially prominent, indicating a pronounced spiral asymmetry. This is qualitatively consistent with the 3D analysis of \citep{walk_neutrino_2020}, in which dipolar and quadrupolar SASI activity dominate in different phases and leave clear multimessenger imprints. In our case, the earliest post-bounce asymmetry may still contain some imprint of the Cartesian grid, so we interpret the very early mode growth with caution; however, the later low-order $\ell=1$ and $\ell=2$ structure in B0R1 is consistent with rotationally modified SASI activity rather than a purely numerical artifact. Such low-order spiral and quadrupolar shock deformations are also of interest for multimessenger diagnostics, since they are expected to modulate both the neutrino signal and the characteristic time-frequency structure of the gravitational-wave emission \citep{walk_neutrino_2020}. In B0R1, these deformations, together with centrifugal support, are associated with a slower but more spatially extended early shock expansion, yielding a larger average shock radius than in B0R0 at this stage. By contrast, B1R1 does not show comparable long-term SASI growth.

By $t\approx 300$\,ms, the models have clearly separated into three distinct flow geometries: B1R1 shows a strongly axis-aligned bipolar structure, B0R1 shows a smaller-scale, more irregular and less strongly directed outflow, and B0R0 exhibits large, turbulent high-entropy plumes without a preferred polar orientation. The rapid shock radius expansion of B0R0 at this time overtakes the more slowly expanding B0R1. These trends persist to $t\approx 500$\,ms, when B1R1 retains the clearest bipolar morphology, B0R1 remains the most compact of the three, and B0R0 shows the broadest and most disordered high-entropy ejecta.

\begin{figure}[t]
    \centering
    \includegraphics[width=\columnwidth]{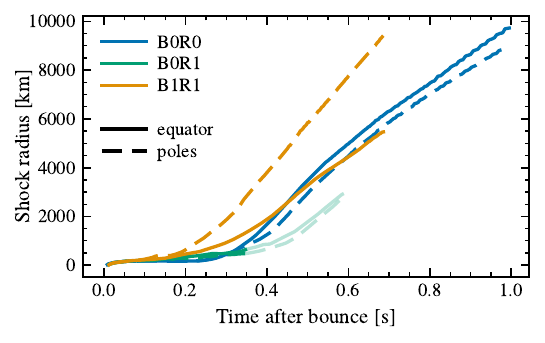}
    \caption{Evolution of the equatorial and polar shock radii as a function of time after core bounce for models B0R0, B0R1, and B1R1. Solid curves show the equatorial shock radii and dashed curves show the polar shock radii. Faded segments indicate the late-time portion of B0R1 affected by the neutrino-transport issue discussed in Sect.~\ref{subsec:resolution}; this interval is shown for completeness but should be interpreted with caution. B1R1 develops an early polar-dominated expansion, whereas B0R0 expands without a strong polar--equatorial asymmetry. B0R1 exhibits the slowest shock expansion over the time interval shown.}
    \label{fig:shock_radii}
\end{figure}

These morphological differences are consistent with the shock-radius evolution and spherical harmonic decomposition shown in Fig.~\ref{fig:shock_radii} and Fig.\ref{fig:sasi_modes}, respectively. B1R1 reaches sustained expansion first, at $t_{\rm rev}\approx 150$\,ms, and is the only model to develop a strong separation between polar and equatorial shock expansion. B0R0 follows at $t_{\rm rev}\approx 250$\,ms and expands in a much less ordered, plume-dominated fashion, although its equatorial and polar radii remain similar. Such a morphology is common in non-rotating 3D simulations, where continued accretion through narrow downflow channels sustains neutrino heating and drives shock expansion in one or more large-solid-angle plumes \citep{burrows_overarching_2020, bollig_self-consistent_2021}. B0R1 reaches sustained expansion latest, at $t_{\rm rev}\approx 370$\,ms, consistent with its comparatively gradual and compact evolution in Fig.~\ref{fig:slices}.

We quantify the energetic growth of the outflow using the diagnostic energy, defined as
\begin{equation}
E_{\rm diag}(t) \equiv \int_{\mathcal{U}(t)} \rho\, e_{\rm tot}\, dV,
\end{equation}
where
\begin{equation}
e_{\rm tot} = e_{\rm int} + \frac{1}{2}v^2 + \frac{B^2}{2\rho} + \Phi ,
\end{equation}
i.e., the sum of the specific internal, kinetic, magnetic, and gravitational potential energies.
The integral is taken over outward-moving unbound material, defined here by $v_r>0$ and positive total specific energy, $e_{\rm tot}>0$. To remove the dependence on the nuclear EOS reference mass in the internal energy, we evaluate the internal-energy contribution relative to cold matter at the same density and $Y_e$. As a caveat, the reported diagnostic energy does not account for the overburden of material at larger radii, and thus should not be interpreted as the final asymptotic diagnostic energy. Determining the final diagnostic energy requires self-consistent evolution to much later times. 

\begin{figure}[t]
    \centering
    \includegraphics[width=\columnwidth]{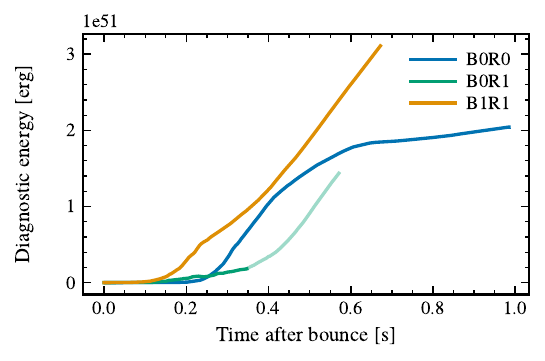}
    \caption{Diagnostic energy, $E_{\rm diag}$, as a function of time after bounce for models B0R0, B0R1, and B1R1. The magnetized rotating model B1R1 shows the earliest and most rapid rise, while B0R0 turns on later and continues increasing through the end of the simulation. Model B0R1 begins its rise at even later times, but once underway its growth rate is comparable to B1R1 over the interval where both are available.}

    \label{fig:diag_energy}
\end{figure}

Fig.~\ref{fig:diag_energy} shows the evolution of $E_{\rm diag}$ for all three models. B0R0 exhibits a later rise that continues for the remainder of the simulation, but its growth rate decreases at late times as the accretion rate onto the PNS, and therefore the neutrino heating, subsides. B0R1 increases more slowly over the overlapping time interval as the later shock expansion develops. 

B1R1 shows a different evolution. The diagnostic energy begins to rise earliest, consistent with the prompt, magnetically aided launch of the bipolar outflow discussed in more detail in \S~\ref{subsec:launch}. At late times, the continued rise of $E_{\rm diag}$ suggests that the shock propagation is being fueled by an energy reservoir that remains available after the initial shock revival. 
In a magnetized rotating model, the natural candidate is the free energy of differential rotation stored in the PNS. As we show in \S~\ref{sec:angmom}, B1R1 undergoes sustained magnetic redistribution of angular momentum and a corresponding decline of the rotational-free-energy reservoir, while $E_{\rm diag}$ continues to grow over the same interval. Although the present diagnostics do not isolate the detailed conversion pathway, the combined evolution is consistent with continued extraction of rotational free energy by magnetic stresses and its transfer into the expanding outflow. Figs.~\ref{fig:slices}--\ref{fig:diag_energy} show that, although all three models reach sustained shock expansion within the simulated window, rotation and magnetic fields strongly influence the geometry of the expanding region and the rate at which diagnostic energy accumulates.

The shock and outflow evolution is closely tied to the continuing supply of accreting material and therefore to the neutrino luminosities and heating conditions discussed in the next section. After the early post-bounce phase, rotation and magnetic stresses produce increasingly different inner-core configurations in the three models, which in turn affect the neutrino emission and heating conditions. We quantify the neutrino diagnostics first, and return to the PNS structure in more detail below.

\subsection{Neutrino diagnostics}
\label{sec:neutrino}

Neutrino emission and absorption in the post-shock region regulate the conditions for stalled-shock revival in the neutrino-heating scenario and remain a key diagnostic even when additional channels (e.g., magnetic stresses) restructure the flow \citep{bethe_revival_1985}. A detailed angular- and spectral-dependent analysis of the neutrino signal, together with multimessenger implications, is deferred to a companion paper. Here we focus on the baseline quantities most directly connected to the global dynamics discussed in \S~\ref{sec:shock}: the angle-averaged neutrino luminosities and the net neutrino heating rate integrated over the gain region.

\begin{figure*}[t]
\centering
\includegraphics[width=\linewidth]{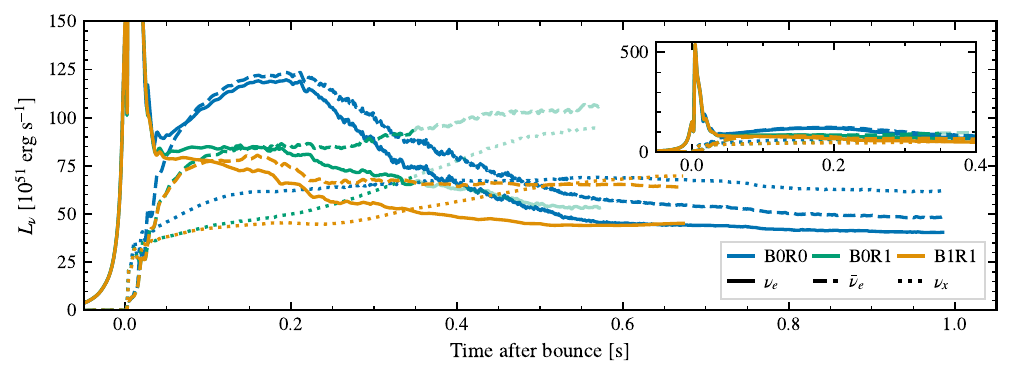}
\caption{Angle-averaged neutrino luminosities, measured on a sphere of radius $500\,\mathrm{km}$, as a function of time after core bounce for models B0R0, B0R1, and B1R1. Solid, dashed, and dotted curves correspond to $\nu_e$, $\bar{\nu}_e$, and $\nu_x$, respectively. The inset highlights the prompt $\nu_e$ burst at shock breakout. Differences at later times largely track changes in the accretion-powered component of the electron-flavour emission as the post-shock flow evolves.}
\label{fig:nu_lum}
\end{figure*}

Fig.~\ref{fig:nu_lum} shows the time evolution of the angle-averaged luminosities for $\nu_e$, $\bar{\nu}_e$, and $\nu_x$ (plotted as the value of a single heavy-lepton species), extracted on a sphere of radius $500\,\mathrm{km}$, for all three models. All models exhibit the canonical prompt $\nu_e$ burst at shock breakout, followed by an accretion-dominated phase in which the electron-flavour luminosities exceed the heavy-lepton luminosity \citep{muller_new_2014}. During $\sim 0.05$--$0.3\,\mathrm{s}$ after bounce, B0R0 reaches the largest electron-flavour luminosities, consistent with its more compact PNS in the absence of rotational support. At later times, the electron-flavour luminosities decline as the accretion-powered contribution to the emission weakens.

The rotating models maintain elevated electron-flavour luminosities to later times. In B0R1, both $\nu_e$ and $\bar{\nu}_e$ luminosities remain relatively high to the end of the simulation. In B1R1, the electron-flavour luminosities decline more rapidly at late times than in B0R1, consistent with the increasing dynamical influence of magnetic stresses and the developing polar outflow. Across all models, the heavy-lepton luminosity varies more slowly and shows a more gradual evolution than the electron flavours, as expected when a significant fraction of the $\nu_x$ emission is set by diffusion from hot dense layers near the PNS \citep{buras_two-dimensional_2006}.

A notable feature of the rotating models is the late-time excess of $L_{\bar{\nu}_e}$ over $L_{\nu_e}$, especially in B0R1. We interpret this as a consequence of the rotationally deformed and extended PNS environment in the rotating simulations: rotation maintains a more oblate PNS, a more extended lower-density envelope, and sustained asymmetric accretion, all of which can modify the electron-flavour decoupling region. \citep{summa_rotation-supported_2018} found the related general effect that rapid rotation lowers the electron-flavour luminosities by producing a more extended and cooler PNS, though without a comparably strong inversion. More directly, \citep{bugli_three-dimensional_2023} reported $L_{\nu_e}<L_{\bar{\nu}_e}$ in rapidly rotating magnetized 3D models and linked it to a larger $\nu_e$--$\bar{\nu}_e$ neutrinosphere separation and to more neutron-rich material around the PNS. Likewise, \citep{obergaulinger_magnetorotational_2021} showed strongly asymmetric electron-flavour emission in rapidly rotating magnetized 3D models, with substantial late-time $\bar{\nu}_e$ excesses in some cases. The particularly strong inversion in B0R1 should be treated with caution given the late-time transport issues in that model. Nevertheless, the same qualitative trend, though at a more modest level, is also present in B1R1 and is broadly consistent with previous rapidly rotating magnetized 3D simulations.

\begin{figure}[t]
\centering
\includegraphics[width=\linewidth]{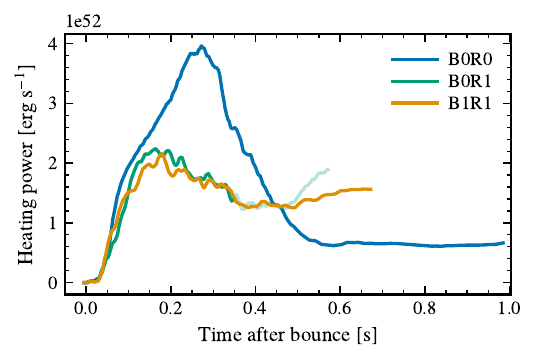}
\caption{Net neutrino heating rate, $\dot{Q}_{\mathrm{heat}}$, integrated over the gain region as a function of time after core bounce for models B0R0, B0R1, and B1R1. All models show a rapid early rise in heating, followed by model-dependent evolution as rotation and magnetic stresses modify the accretion flow and the structure of the gain layer.}
\label{fig:nu_heating}
\end{figure}

To connect the luminosity evolution to the shock expansion conditions, Fig.~\ref{fig:nu_heating} shows the net neutrino heating rate, $\dot{Q}_{\mathrm{heat}}$, integrated over the gain region. All three models exhibit a rapid rise in $\dot{Q}_{\mathrm{heat}}$ during the early post-bounce phase, but B0R0 reaches the largest values, peaking at $\sim 4\times10^{52}\ \mathrm{erg\ s^{-1}}$ at $\sim 0.25\,\mathrm{s}$. This is consistent with the standard heating -- advection picture of shock revival, in which stronger neutrino heating and longer dwell times in the gain region favour shock expansion \citep{janka_conditions_2001,janka_explosion_2012,murphy_criteria_2008,couch_revival_2013,boccioli_neutrino_2025}. After this peak, $\dot{Q}_{\mathrm{heat}}$ declines steadily in B0R0 as the gain-layer structure and accretion flow evolve.

The rotating models remain at lower heating rates overall. B1R1 stays relatively flat at late times, settling to $\sim (1.3$--$1.6)\times10^{52}\ \mathrm{erg\ s^{-1}}$ over $\sim 0.4$--$0.7\,\mathrm{s}$. B0R1 also remains below B0R0, but its apparent late-time rise toward $\sim 2\times10^{52}\ \mathrm{erg\ s^{-1}}$ by $\sim 0.6\,\mathrm{s}$ should be interpreted with caution, since the final phase of this model shows numerically suspicious behaviour (see \S~\ref{subsec:resolution}) and this increase may not be entirely physical. Overall, the reduced heating in the rotating models is consistent with the lower electron-flavour luminosities and more extended PNS structure discussed above \citep{janka_conditions_2001,summa_rotation-supported_2018}. In B1R1, magnetic stresses additionally help drive polar expansion and reorganize the accretion flow (see \S~\ref{sec:jets}), so sustained shock expansion does not require the comparatively large gain-region heating rates reached in B0R0.

More detailed neutrino diagnostics, including angular dependence and observer-dependent signals, will be presented in a forthcoming companion study.

\subsection{Proto-neutron star structure and rotational reservoir.}
\label{sec:pns}

The PNS is the central engine of the post-bounce evolution: it sets the neutrino emission through its thermal and accretion-powered luminosities and, in rotating models, stores a reservoir of rotational free energy that can be tapped by magnetic stresses. In this section we summarize the PNS mass growth, geometric evolution (oblateness), and an estimate of the available rotational free energy in the rotating models.

\subsubsection{Proto-neutron star mass growth.}
\label{subsec:pns_mass}
\begin{figure}[t]
    \centering
    \includegraphics[width=\columnwidth]{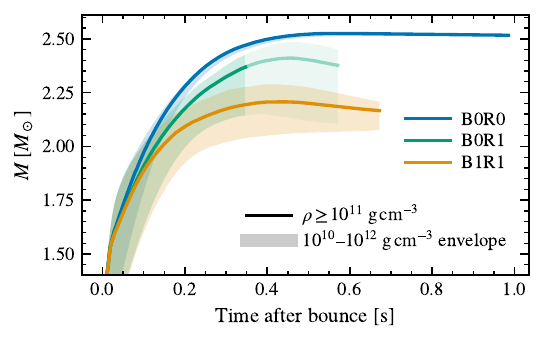}
    \caption{Time evolution of the enclosed \emph{baryonic} PNS mass for models B0R0, B0R1, and B1R1. The solid curves show the mass enclosed above a density threshold of $\rho \geq 10^{11}\,\mathrm{g\,cm^{-3}}$, while the shaded bands span density thresholds from $10^{10}$ to $10^{12}\,\mathrm{g\,cm^{-3}}$. All models show rapid growth during the early post-bounce accretion phase, followed by a slower increase and approximate saturation by $\sim 0.4$--$0.5\,\mathrm{s}$. The enclosed mass remains largest in B0R0 and smallest in B1R1 at late times. Unlike B0R0, the rotating models retain a significant fraction of their mass at lower densities owing to rotational support. In B1R1, magnetic stresses further redistribute material and contribute to keeping more of it at lower densities, thereby reducing the enclosed mass defined by the chosen PNS density threshold.}
    \label{fig:pns_mass}
\end{figure}

Fig.~\ref{fig:pns_mass} shows the evolution of the enclosed baryonic PNS mass for all three models, where the PNS is defined by a density threshold of $\rho \ge 10^{11}\,\mathrm{g\,cm^{-3}}$. In each case the PNS mass increases rapidly during the early post-bounce accretion phase and then approaches a plateau by $\sim 0.4$--$0.5\,\mathrm{s}$. The final baryonic masses follow a clear ordering: the non-rotating hydrodynamic model B0R0 reaches the largest value ($\simeq 2.5\,M_\odot$), the rotating non-magnetized model B0R1 is intermediate ($\simeq 2.4\,M_\odot$), and the magnetized rotating model B1R1 is smallest ($\simeq 2.2\,M_\odot$). This ordering is broadly consistent with the differing flow morphologies, but it should be interpreted with some caution because the PNS mass depends on the density threshold, especially for rotating models. In particular, at early times all three models retain a non-negligible amount of matter immediately outside the $\rho=10^{11}\,\mathrm{g\,cm^{-3}}$ surface: about $\sim 10^{-1}\,M_\odot$ residing in the interval $10^{10}\lesssim \rho <10^{11}\,\mathrm{g\,cm^{-3}}$ and a somewhat larger mass of the same order is present at $10^{9}\lesssim \rho <10^{10}\,\mathrm{g\,cm^{-3}}$. In B0R0 this lower-density envelope contracts and is largely incorporated into the dense core, whereas in the rotating models a more substantial fraction remains distributed at lower densities for much longer after the shock expansion sets in. Thus, the smaller enclosed masses in B0R1 and B1R1 do not simply imply that this mass is absent; rather, part of it resides in a more extended, lower-density envelope. In B1R1, magnetically driven outflows likely contribute further by helping to redistribute, and in some cases expel, material from the immediate PNS vicinity.

At late times, both rotating models show a modest decline in the enclosed mass. We interpret this as arising from a combination of redistribution of material across the adopted $10^{11}\,\mathrm{g\,cm^{-3}}$ boundary and some genuine mass loss from the immediate PNS region. In B1R1 this is aided by magnetic torques and the emerging outflow, while in B0R1 late-time low-$T/W$ activity and rotational support help maintain a more extended, lower-density structure. Since the inferred PNS mass depends on the density cutoff, the mass evolution alone does not uniquely disentangle these contributions.

\subsubsection{Proto-neutron star radii and oblateness.}
\label{subsec:pns_radii}
\begin{figure}[t]
    \centering
    \includegraphics[width=\columnwidth]{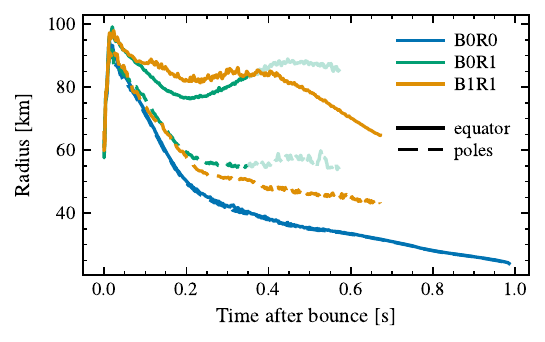}
    \caption{Evolution of the PNS radius measured along the poles (dashed) and in the equatorial plane (solid) for models B0R0, B0R1, and B1R1. The non-rotating model remains nearly spherical, while rotation produces a sustained oblateness with larger equatorial radii by redistributing matter outward and supporting a more extended equatorial structure. In B1R1, magnetic stresses and the emerging polar outflow further modify the inner mass distribution, leading to a smaller late-time equatorial radius than in B0R1.
    }
    \label{fig:pns_radii}
\end{figure}

Fig.~\ref{fig:pns_radii} shows the evolution of the PNS radii measured along the poles and in the equatorial plane. All models exhibit an initial rapid contraction after bounce as the PNS deleptonizes and cools. The subsequent evolution differs between rotating and non-rotating cases. In B0R0 the contraction proceeds relatively uniformly, yielding nearly identical polar and equatorial radii throughout the evolution. In the rotating models, centrifugal support produces a pronounced oblateness: equatorial radii remain substantially larger than polar radii. This flattening is strongest in the non-magnetized rotating model B0R1, while B1R1 maintains smaller radii at late times, as magnetic angular-momentum transport reduces centrifugal support and allows the PNS to contract more efficiently. 

\subsubsection{Rotational free energy.}
\label{subsec:pns_frot}
\begin{figure}[t]
    \centering
    \includegraphics[width=\linewidth]{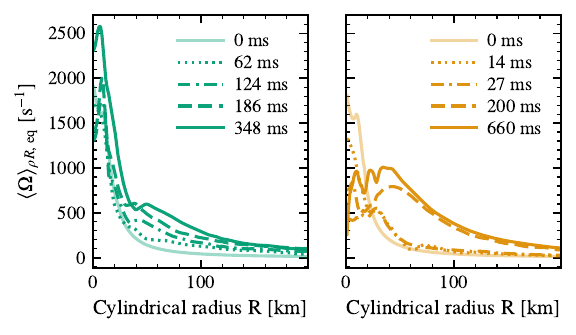}
    \caption{Equatorial, density-weighted, averaged angular-velocity profiles, $\langle\Omega\rangle_{\rho}(R)$, as a function of cylindrical radius $R$ for models B0R1 and B1R1 at selected post-bounce times. B0R1 retains a strongly differential and centrally peaked rotation profile throughout the evolution, whereas B1R1 develops a broader and flatter inner profile, indicating more efficient redistribution of angular momentum by magnetic stresses.}
    \label{fig:omega_profiles}
\end{figure}

In the rotating models, differential rotation in the newly formed PNS stores free energy that can in principle be extracted by magnetic stresses. In B1R1, magnetic amplification and the resulting Maxwell stresses transport angular momentum outward, thereby reducing the inner shear and tapping this reservoir, consistent with previous rapidly rotating magnetized 3D simulations \citep{bugli_three-dimensional_2023}. This redistribution is also visible in the equatorial, density-weighted angular-velocity profiles shown in Fig.~\ref{fig:omega_profiles}: B0R1 retains a more sharply peaked and strongly differential inner rotation profile, whereas B1R1 develops a broader and flatter profile. Figure~\ref{fig:omega_profiles} therefore gives a spatial view of the same spin-down process inferred from the rotational-energy evolution: magnetic stresses in B1R1 reduce the central shear and spread angular momentum outward, while B0R1 preserves a more compact differentially rotating core.

\begin{figure}[t]
    \centering
    \includegraphics[width=\columnwidth]{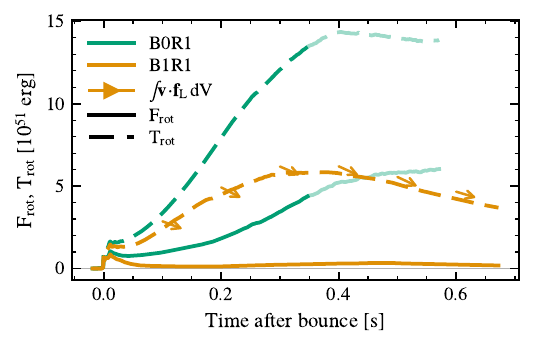}
    \caption{Evolution of the rotational free energy, $F_{\rm rot}$ (solid), and total rotational energy, $T_{\rm rot}$ (dashed), for the rotating models B0R1 and B1R1. Arrows along the B1R1 curve mark the Lorentz-work contribution in the PNS, illustrating the extraction of rotational energy by magnetic stresses. Compared to B0R1, the magnetized model settles to a smaller free-energy reservoir after the early post-bounce phase, consistent with magnetic extraction of rotational free energy during the growth of the polar outflow. All quantities are computed over the PNS.}
    \label{fig:frot}
\end{figure}

Following the convention of \citep{dessart_arduous_2012}, we estimate the available reservoir by comparing the instantaneous rotational energy of the differentially rotating PNS to that of an equivalent uniformly rotating configuration with the same total angular momentum and moment of inertia. We define
\begin{equation}
F_{\rm rot} \equiv T - \frac{1}{2} I \,\bar{\Omega}^{\,2}
\;=\;
T - \frac{J^{2}}{2I},
\label{eq:frot_def}
\end{equation}
where $T$ is the rotational kinetic energy of the PNS, $J$ is its total angular momentum about the rotation axis, $I$ is its moment of inertia, and $\bar{\Omega}\equiv J/I$ is the angular velocity of the corresponding uniformly rotating configuration. Because this estimate depends explicitly on the PNS moment of inertia, we assess the resolution sensitivity of $I$ in Appendix~\ref{app:moi_resolution}. Rather than focusing on $F_{\rm rot}$ alone, Fig.~\ref{fig:frot} shows the evolution of both the rotational free energy, $F_{\rm rot}$, and the total rotational energy, $T_{\rm rot}$, for the two rotating models.

The two rotating models exhibit clearly different evolutions of this quantity. In the non-magnetized rotating model B0R1, both the rotational free energy $F_{\rm rot}$ and the total rotational energy $T_{\rm rot}$ grow during the early post-bounce phase and remain comparatively high thereafter, which indicates that a substantial reservoir of differential rotational energy is maintained throughout the evolution and is untapped. 

In the magnetized model B1R1, $T_{\rm rot}$ follows a broadly similar early-time rise but saturates at a lower level than in B0R1, while $F_{\rm rot}$ remains consistently smaller at all times. This behaviour indicates that, although differential rotation is still present, the available rotational free-energy reservoir is reduced relative to the non-magnetized case. The difference is consistent with the magnetic field tapping this reservoir: magnetic winding amplifies the field, and the resulting Maxwell stresses redistribute angular momentum and extract energy from differential rotation. The quantity $\dot{W}_{L}^{\rm PNS}$ provides a direct measure of this coupling between the magnetic field and the fluid inside the PNS. For Fig.~\ref{fig:frot} we overplot the instantaneous Lorentz-work rate integrated over the PNS,
\begin{equation}
\dot{W}_{L}^{\rm PNS} \equiv \int_{\rm PNS} \mathbf{v}\!\cdot\!\mathbf{f}_{L}\, dV,
\label{eq:lorentz_work_pns}
\end{equation}
which measures the rate at which magnetic stresses do mechanical work on the fluid within the PNS. The physical mechanisms responsible for this extraction, including magnetic winding and Maxwell stresses, are examined in more detail in \S~\ref{sec:angmom}, while the late-time energetic consequences of this extraction are discussed in \S~\ref{subsec:magbudget}.

In summary, the PNS properties reflect the different accretion and outflow geometries established in \S~\ref{sec:shock} and are directly connected to the neutrino and magnetic channels discussed in subsequent sections. The non-rotating model forms the most massive and most compact PNS and undergoes collapse to a black hole by the end of the simulated interval, while the rotating models maintain larger equatorial radii due to centrifugal support and retain a significant reservoir of rotational free energy.

\subsection{Angular-momentum transport and magnetic spin-down}
\label{sec:angmom}

Rotation supplies the PNS with a substantial reservoir of rotational free energy, but the subsequent evolution depends on whether that reservoir can be redistributed within the PNS or extracted into the outflow. In the rotating non-magnetized model B0R1, angular momentum supplied by continued accretion is largely retained within the PNS, allowing strong differential rotation and centrifugal support to persist. In the magnetized model B1R1, by contrast, magnetic winding and the resulting Maxwell stresses provide an additional transport channel that redistributes angular momentum and extracts rotational energy from the inner core. The resulting magnetic spin-down is one of the clearest dynamical differences between the two rotating models.

We quantify this difference with a control-volume angular-momentum budget evaluated within a spherical surface of radius $R=50\,\mathrm{km}$. As in the flux-based angular-momentum accounting of \citet{powell_three_2023}, we focus on the component along the initial rotation axis, $J_y$, but here we also include the neutrino-radiation contribution. The corresponding source terms of the angular-momentum budget for the component along the rotation axis, $J_y$, can be written as

\begin{equation}
\frac{\partial J_y(<R,t)}{\partial t}
=
-\int_{A_R} \rho\,(\mathbf{r}\times\mathbf{v})_y\,v_r\,dA
+\int_{A_R} \frac{1}{4\pi}\,(\mathbf{r}\times\mathbf{B})_y\,B_r\,dA
\end{equation}
\begin{equation}
\qquad
-\sum_{\nu,\epsilon}\int_{A_R}
\left[\mathbf{r}\times\left(\mathbf{P}^{(\nu)}_{\epsilon}\cdot\hat{\mathbf{e}}_r\right)\right]_y\,dA.
\label{eq:jy_budget_full}
\end{equation}
\noindent
where the three terms on the right-hand side represent, respectively, the advective angular-momentum flux, the magnetic torque, and the neutrino-related contribution. The derivation of equation~(\ref{eq:jy_budget_full}) from the ideal-MHD momentum equation together with the neutrino contribution derived from the two-moment radiation field is given in Appendix~\ref{app:torque}.

To interpret the budget more directly, we reconstruct the enclosed angular momentum by integrating Eq.~\ref{eq:jy_budget_full} in time. We repeat this reconstruction while including different combinations of the surface-flux terms. In each case, the surface fluxes are evaluated at 1\,ms intervals. For the neutrino-induced angular-momentum change, however, the available post-processing output is temporally sparse, so this contribution is interpolated to the 1\,ms sampling used in the reconstruction. As in other multidimensional CCSN calculations, such reconstructions should not be expected to close perfectly: deviations can arise both from genuine redistribution of angular momentum across the control-volume boundary and from imperfect numerical conservation or diffusion in the underlying hydrodynamics \citep{cabezon_core-collapse_2018}.

\begin{figure}
\centering
\includegraphics[width=\linewidth]{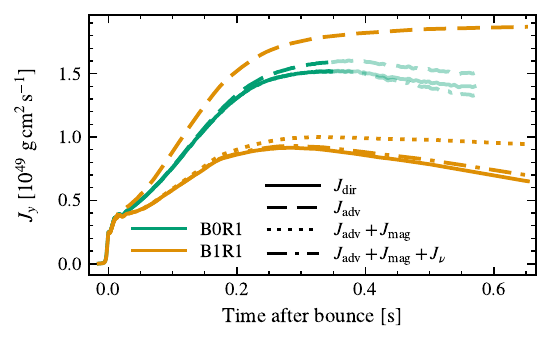}
\caption{Evolution of the angular momentum $J_y$ enclosed within a sphere of radius $R=50\,\mathrm{km}$ in the rotating models B0R1 and B1R1. Solid lines show the directly measured enclosed angular momentum, $J_{\mathrm{dir}}$. Dashed lines show the cumulative advective contribution, dotted lines show the cumulative advective plus magnetic contribution (for B1R1 only), and dash-dotted lines show the cumulative advective plus magnetic (the magnetic contribution being zero for B0R1) plus neutrino-related momentum loss. In both models the advective contribution is positive and grows rapidly after bounce, indicating continued inward supply of positive $J_y$. In B1R1, however, the direct curve lies far below the purely advective expectation, and only approaches the reconstructed evolution once magnetic losses are included. The neutrino contribution is smaller than the magnetic term, but provides an additional negative correction that further improves the agreement with the direct evolution, particularly at late times.}
\label{fig:angular_momentum}
\end{figure}

Fig.~\ref{fig:angular_momentum} shows the resulting decomposition. In both rotating models, the cumulative advective term rises steadily after bounce, indicating continued inward delivery of positive angular momentum through the $50\,\mathrm{km}$ surface. Notably, this cumulative advective contribution is larger in B1R1 than in B0R1, even though the magnetized model retains less angular momentum inside $50\,\mathrm{km}$. This is because the advective term tracks transport across a fixed surface rather than storage inside it: in B1R1, the stronger PNS contraction relative to B0R1 advects rotating outer PNS material through the fixed $50\,\mathrm{km}$ surface, while magnetic stresses prevent the associated positive $J_y$ from accumulating in the inner core.

\begin{figure}
\includegraphics[width=\linewidth]{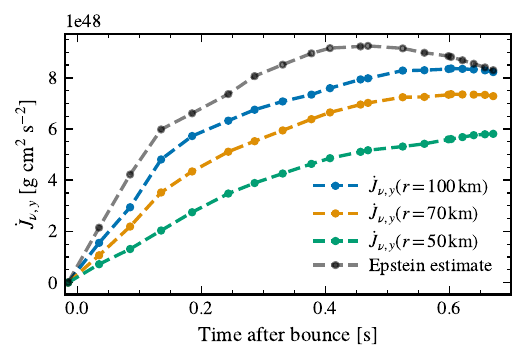}
\caption{Comparison of the neutrino-associated angular-momentum emission rate in the rotating magnetised model B1R1 to the analytic estimate of \citet{epstein_neutrino_1978}. The coloured curves show the directly measured $\dot J_{y,\nu}$ through spherical surfaces at radii of $50$, $70$, and $100\,\mathrm{km}$. The gray dashed curve shows the diffusion-limit Epstein estimate, Eq.~\ref{eq:epstein}. Filled circles mark the simulation outputs at which the neutrino angular-momentum flux was evaluated directly.}
\label{fig:jy_nu}
\end{figure}

In B0R1, the directly measured enclosed angular momentum remains comparatively close to this advective expectation, showing that most of the delivered $J_y$ is retained within the inner PNS. In B1R1, however, the direct evolution differs strongly from the purely advective reconstruction: advection alone substantially overpredicts the retained angular momentum. Including the magnetic term removes most of this discrepancy, demonstrating that Maxwell stresses provide the dominant additional sink required to recover the measured spin evolution. The neutrino contribution acts in the same direction, but remains smaller over most of the evolution and mainly provides a secondary correction, improving the agreement further at late times. 

Thus, while continued accretion attempts to spin up the inner core in both rotating models, only in the magnetized case is that advective input strongly offset by an efficient outward transport channel. We note that this balance is evaluated at a fixed radius of $50\,\mathrm{km}$; at larger radii, where the PNS continues to restructure over longer timescales, the magnetic contribution is expected to become even more important at late times as the region over which Maxwell stresses operate extends outward.

The neutrino-related term in the budget is smaller than the magnetic torque, but it is still non-negligible and merits separate discussion. To our knowledge, the angular momentum change due to neutrino emission has not been calculated for self-consistent simulations of magnetorotationally driven supernovae before (although see \cite{harada_neutrino_2019} for a calculation in 2D rotating Boltzmann transport simulations) and therefore we briefly elaborate on it. The concept was first proposed by \cite{kazanas_neutrino_1977,epstein_neutrino_1978} as a method of spinning down neutron stars. From \cite{epstein_neutrino_1978}, in the slow rotation limit, the angular momentum emitted by an optically thick (diffusion limit), uniformly spinning neutron star is given by
\begin{equation}
\dot J_{y,\nu}^{\mathrm{Ep}}
=
\frac{2}{3c^2}\,L_\nu\,R^2\,\Omega,
\label{eq:epstein}
\end{equation}
\noindent
where $L_\nu$ is the total neutrino luminosity from the object, $R$ is its radius, and $\Omega$ is the uniform angular velocity. Our neutrino transport is solved with a multidimensional two-moment scheme with an analytic closure. Knowledge of the second moment (or an analytic expression for it) of the neutrino distribution allows us to evaluate the radial flux of tangential momentum, and therefore allows us to determine the flux of angular momentum through any surface, i.e. Eq.\ref{eq:djnudt_sphere}. Fig.~\ref{fig:jy_nu} compares the directly measured neutrino angular-momentum emission rate, evaluated from the two-moment radiation field via Eq.~\ref{eq:djnudt}, at several radii to the Epstein estimate in Eq.~\ref{eq:epstein}. For the latter, as a proxy for the neutrinosphere, we take $R$ and $\Omega$ as the average values near $\rho=10^{11}$\,g\,cm$^{-3}$ and $L_\nu$ as the total luminosity at 500\,km. We show the direct neutrino angular momentum flux evaluated at radii of $50$, $70$, and $100\,\mathrm{km}$ and is plotted as $\dot J_{y,\nu}$, so that positive values correspond to loss of positive angular momentum from the inner region. The neutrino contribution grows steadily after bounce and reaches several $\times 10^{48}$\,g\,cm$^2$\,s$^{-1}$. Generally, as the radius of the sphere increases (and the matter transitions from optically thick to optically thin) there is more net momentum in the neutrino field, increasing the emitted angular momentum.

The Epstein estimate captures the correct order of magnitude, but its limitations are apparent. Eq.~(\ref{eq:epstein}) assumes an idealised, effectively isotropic diffusion picture, whereas in the simulation the neutrino radiation field is neither perfectly spherical nor purely diffusion-like throughout the region of interest. In particular, large scale spatial anisotropies in the neutrino flux due to the rotation as well as the extended decoupling region due to varying neutrino species and energies both modify the angular-momentum flux via neutrinos relative to the simple analytic picture. Nevertheless, this analytic estimate reproduces the approximate scale of the neutrino angular momentum flux reasonably well. At the same time, it generally lies toward the upper end of the directly measured values, especially at smaller radii, where the assumptions entering the analytic estimate are least well satisfied. The agreement improves at larger radii and at later times, when those radii more closely encompass the bulk of the neutrino emission after the PNS has contracted. For this reason, we use the Epstein estimate only as a comparison diagnostic; the neutrino angular-momentum flux entering the cumulative budget in Fig.~\ref{fig:angular_momentum} is taken directly from the two-moment radiation field.

Although Fig.~\ref{fig:angular_momentum} focuses on the component aligned with the initial rotation axis, the rotating models also develop non-zero angular-momentum components, $J_x$ and $J_z$, indicating that the angular momentum supplied by post-bounce accretion is not perfectly aligned with $\hat{\mathbf{y}}$. In B1R1 these components remain subdominant compared to $J_y$, but they still become dynamically relevant for the modest late-time tilt of the PNS spin axis discussed in §~\ref{subsec:tilt}. The diagnostics further suggest that the off-axis angular momentum is first deposited by asymmetric accretion in the outer PNS, after which magnetic torques of the opposite sign act to redistribute and partially remove it. Thus, the transverse angular momentum does not simply accumulate advectively: accretion introduces it, and the magnetic field responds by mediating its subsequent redistribution.

Figs.~\ref{fig:angular_momentum} and \ref{fig:jy_nu} show that the post-bounce evolution of the inner-core spin is controlled by an interplay of inward advective supply and outward removal by magnetic stresses and neutrino emission. In B0R1 the retained angular momentum remains comparatively close to the cumulative advective expectation, in B1R1, by contrast, magnetic torques substantially offset the positive advective influx, while neutrino losses provide an additional, smaller sink for both. The result is a significantly reduced retained $J_y$ in the magnetised model, despite continued inward advection of positive angular momentum through the $50\,\mathrm{km}$ surface. This strong reduction of the retained inner-core angular momentum in B1R1 is broadly consistent with the rapid magnetic spin-down found in previous 3D magnetorotational CCSN simulations of rapidly rotating progenitors \citep{powell_three_2023, obergaulinger_magnetorotational_2021}.

\subsection{Magnetically driven polar outflow: launch, energetics, and 3D stability}
\label{sec:jets}

We now focus on the magnetized rotating model B1R1, which develops the earliest magnetically aided bipolar outflow and the clearest jet-like polar morphology (\S~\ref{sec:shock}). In contrast, the non-rotating hydrodynamic model B0R0 remains comparatively spherical, while the rotating non-magnetized model B0R1 does not develop a comparably clear bipolar outflow. Instead, at later times B0R1 shows weaker, less collimated high-entropy polar features in the entropy slices of Fig.~\ref{fig:slices}, but its overall shock expansion remains less strongly organized into a persistent jet-like morphology than in B1R1. The evolution of B1R1 is marked by the rapid formation of a magnetically dominated polar funnel that subsequently launches a sustained bipolar outflow, a behaviour qualitatively consistent with previous magnetorotational core-collapse supernova simulations \citep{burrows_simulations_2007,winteler_magnetorotationally_2012,mosta_magnetorotational_2014,kuroda_magnetorotational_2020,obergaulinger_magnetorotational_2021}. A time-resolved volume rendering of the specific magnetic-energy density is provided in the online supplementary material and illustrates the three-dimensional development of the magnetized outflow throughout the simulation.

\subsubsection{Launching mechanism}
\label{subsec:launch}

\begin{figure*}[t]
    \centering
    \includegraphics[width=1\linewidth]{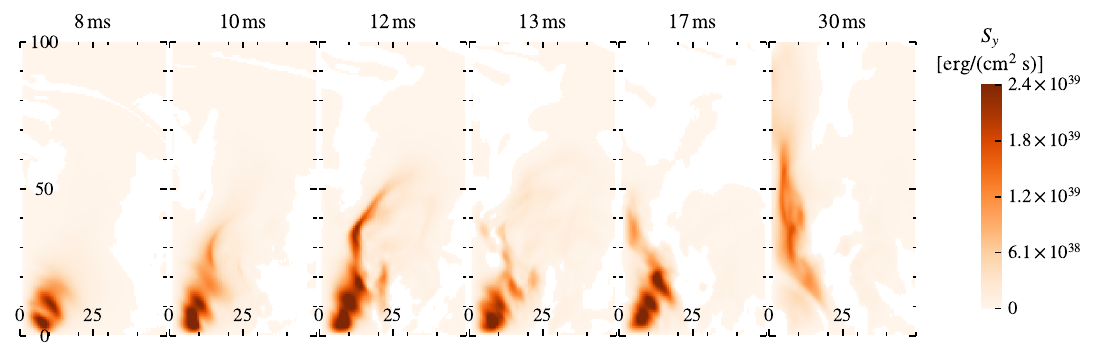}
    \caption{Time sequence of the northward Poynting flux $S_y$ in model B1R1, shown in azimuthally averaged (around the rotation axis) meridional slices constructed from a $200\times200\times200\,\mathrm{km}$ cube centered on the PNS, with only the northern half-plane displayed, at $t_{\rm pb}=8$, 10, 12, 13, 17, and $30\,\mathrm{ms}$. The early emission is broad and poorly collimated, corresponding to an initial transient outflow that weakens rapidly and does not propagate to large radii. It is followed by the emergence of a more centrally concentrated and collimated polar channel, visible by $\sim 13$--$17\,\mathrm{ms}$, which strengthens and extends outward by $30\,\mathrm{ms}$, indicating the onset of a sustained magnetically driven outflow.}
    \label{fig:B1R1_Sy_north}
\end{figure*}

The bipolar outflow is ultimately powered by the free energy of differential rotation in the PNS, tapped by magnetic winding and Maxwell stresses, and redirected into the polar regions. In B1R1, this reservoir is rapidly depleted after bounce (Fig.~\ref{fig:frot}), consistent with efficient magnetic extraction of differential rotational energy. The clearest launch diagnostic is the outward Poynting flux. Fig.~\ref{fig:B1R1_Sy_north} shows an early-time sequence of the azimuthally averaged northward Poynting flux in the meridional plane. The first positive episode appears within the first $\sim 15\,\mathrm{ms}$ after bounce, but it is broad, poorly collimated, and weakens rapidly. This early transient likely reflects the release of an initial magnetic reservoir rather than an efficiently sustained, winding-driven outflow.

\begin{figure}[t]
    \centering
    \includegraphics[width=\columnwidth]{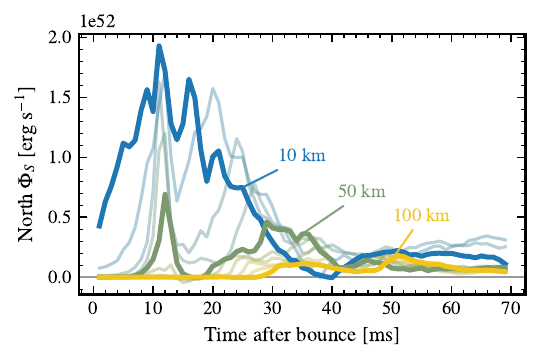}
    \caption{Early-time outward Poynting flux through circular planes of radius $R=30$ km in the north polar direction of model B1R1, measured at heights $y=10$--$100$ km. The first strong positive pulse appears at the smallest sampled heights and subsequently at larger heights, consistent with outward propagation of magnetic energy from the inner PNS/shear-layer system into the polar channel.}
    \label{fig:poynting_flux}
\end{figure}

A second, more coherent outflow episode follows. In Fig.~\ref{fig:B1R1_Sy_north}, this later episode appears as a more centrally concentrated polar channel by $t_{\rm pb}\simeq 13$--$17\,\mathrm{ms}$ and strengthens by $t_{\rm pb}\simeq 30\,\mathrm{ms}$. To quantify its outward propagation, Fig.~\ref{fig:poynting_flux} shows the Poynting flux through fixed circular planes in the north polar direction. The first transient is strongest at the lowest sampled heights and weakens before becoming a sustained large-scale outflow, whereas the later episode produces a more persistent outward flux that reaches progressively larger heights. The sustained outward Poynting flux therefore marks the transition from a failed transient launch to a coherent magnetically powered outflow that can couple the inner PNS region to the stalled shock.

\begin{figure*}[t]
    \centering
    \includegraphics[width=0.75\linewidth]{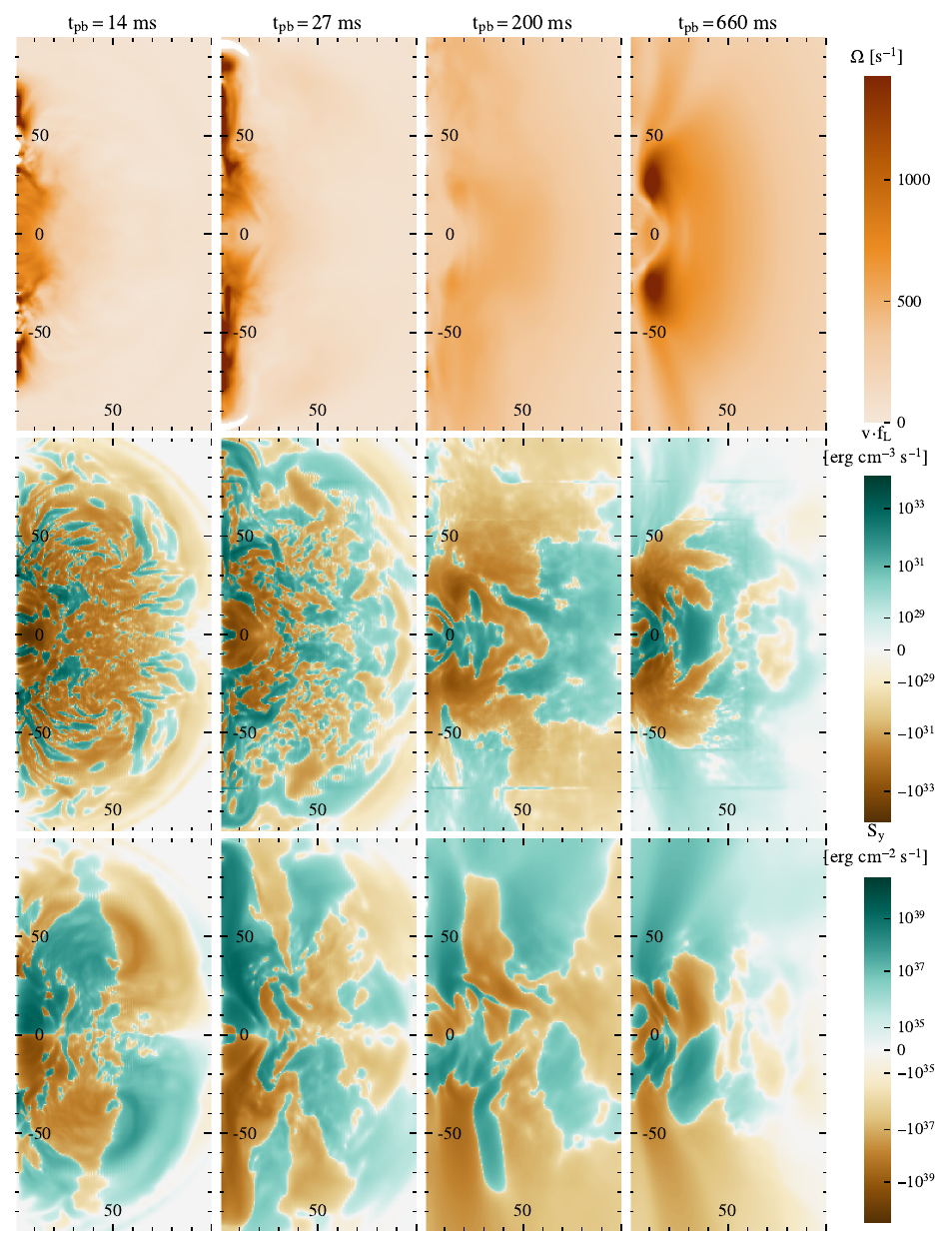}
    \caption{Azimuthally averaged about the rotation axis slices for model B1R1 at four post-bounce times, \(t_{\rm pb}=14\), 27, 200, and 660\,ms (left to right). Each panel shows the inner $200\times200\,\mathrm{km}$ region of the azimuthally averaged meridional plane, constructed from the same $200\times200\times200\,\mathrm{km}$ cube centered on the PNS, with cylindrical radius increasing to the right. The top row shows the angular velocity, \(\Omega\); the middle row shows the local power density \(\mathbf{v}\!\cdot\!\mathbf{f}_L\), where \(\mathbf{f}_L\) is the Lorentz force; and the bottom row shows the \(y\)-component of the Poynting flux, \(S_y\). The two leftmost columns show instantaneous snapshots, while the two rightmost columns are additionally averaged over a 20\,ms window (\(t_{\rm pb}\pm10\)\,ms) to emphasize the larger-scale structure. This averaging helps suppress short-timescale variability and highlights the underlying large-scale structure of the time-dependent, off-axis outflow. Positive and negative values in \(\mathbf{v}\!\cdot\!\mathbf{f}_L\) indicate regions where magnetic forces do positive or negative work on the fluid, respectively, while the sign of \(S_y\) traces the direction of electromagnetic energy transport along the rotation axis. Visualization artifacts occur in the local power density at/near mesh refinement boundaries due to the post processing of gradients across the boundary. An online supplementary movie is provided showing the time evolution of these same azimuthally averaged quantities.}
    \label{fig:B1R1_triptych_axi}
\end{figure*}

This evolution is illustrated in Fig.~\ref{fig:B1R1_triptych_axi}, which shows azimuthally averaged meridional maps of model B1R1, constructed from a $200\times200\times200\,\mathrm{km}$ cube centered on the PNS and displayed $100\times200\,\mathrm{km}$ half-plane at $t_{\rm pb}=14$, $27$, $200$, and $660\,\mathrm{ms}$, including the angular velocity, Lorentz-work density, and vertical Poynting flux. By $t_{\rm pb}=27$\,ms the magnetized outflow extends to the edge of the plotted domain ($\sim 100$ km; Fig.~\ref{fig:B1R1_triptych_axi}$)$, and the later panels at $t_{\rm pb}=200$ and $660$\,ms show that this narrower bipolar outflow persists for the remainder of the simulation. A time-resolved version of the same azimuthally averaged quantities is provided in the online supplementary material.

The magnetic energy within the PNS rises rapidly after bounce, with the toroidal component amplifying especially quickly and dominating the early growth as differential rotation winds the field. This rapid winding builds magnetic pressure, and the resulting axial magnetic-pressure gradient inflates and accelerates the polar flow. The fluid is not mechanically accelerated by the Poynting flux itself; rather, the Poynting flux measures the outward transport of energy, while the Lorentz force accelerates the matter. The magnetic stresses therefore do two things simultaneously: they extract and redistribute angular momentum in the dense rotating core, and they do positive work in the emerging polar outflow. In Fig.~\ref{fig:B1R1_triptych_axi}, positive $\mathbf{v}\cdot\mathbf{f}_{L}$ is concentrated in the polar channel, while $S_y$ shows the direction of the electromagnetic energy flux along the outflow.

After the launch phase, the magnetic structure does not return to a broad, disordered configuration. Instead, Fig.~\ref{fig:B1R1_triptych_axi} shows that the strongest differential rotation becomes concentrated in off-equatorial belts above and below the equatorial plane. These rapidly rotating layers provide the natural sites for continued winding of the poloidal field into a toroidal component, and they coincide with structured Lorentz work and vertical Poynting flux around the PNS. This figure identifies the rotating regions that sustain the wound-field configuration, rather than showing the magnetic energy directly. Over the same interval, the rapid post-bounce growth of the magnetic-energy reservoir slows: both the poloidal and toroidal components evolve more gradually, indicating that the system has moved from rapid field amplification into a more slowly varying, quasi-saturated state.

The later-time azimuthally averaged slices then show how this organized magnetic structure propagates outward. The outflow remains identifiable as a magnetized polar funnel rather than reverting to a broad, disordered configuration. Its fully three-dimensional stability and time-dependent axis motion are quantified separately in Sect.~\ref{subsec:kink}, where we track the lateral displacement and tilt of the funnel.

\subsubsection{Magnetic-energy budget and effective dissipation}
\label{subsec:magbudget}

To connect the launch diagnostics to the later-time energetics, we examine the magnetic-energy budget in a fixed spherical control volume. For a sphere of radius \(R=90\,\mathrm{km}\), we define the enclosed magnetic energy

\begin{equation}
E_B(<R,t) \equiv \int_{V(R)} e_B\,dV,
\qquad
e_B \equiv \frac{B^2}{8\pi}.
\end{equation}
The corresponding volume-integrated magnetic-energy equation can be written as
\begin{equation}
\frac{dE_B}{dt}
=
-\int_{V(R)} \mathbf{v}\!\cdot\!\mathbf{f}_L\,dV
 -\int_{V(R)} \nabla\!\cdot\!\mathbf{S}\,dV
-\dot{Q}_{\rm diss},
\label{eq:mag_budget_main}
\end{equation}
with the local residual \(Q_{\rm diss}\) defined as part of the derivation in Appendix~\ref{app:diss_deriv}. As shown there, \(Q_{\rm diss}\) is the part of the magnetic-energy loss not accounted for by Poynting-flux transport or reversible Lorentz work; in a resistive system it reduces to the Ohmic heating rate, while in ideal MHD any non-zero measured value reflects effective dissipation associated with numerically regulated reconnection and unresolved sub-grid magnetic-energy transfer.

\begin{figure}[t]
    \centering
    \includegraphics[width=\columnwidth]{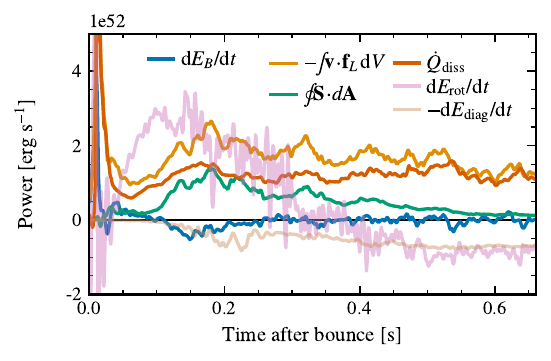}
    \caption{Magnetic power budget for model B1R1 inside a spherical control volume of radius \(R=90\,\mathrm{km}\). Shown are the magnetic-energy derivative \(dE_B/dt\), the volume-integrated Lorentz-work term \(-\int \mathbf{v}\!\cdot\!\mathbf{f}_L\,dV\), the outward Poynting flux through the control surface \(\oint \mathbf{S}\!\cdot\! d\mathbf{A}\), and the residual \(\dot{Q}_{\rm diss}\) required to close the discrete balance in Eq.~\ref{eq:mag_budget_main}. For comparison, we also show \(dE_{\rm rot}/dt\) and \(-dE_{\rm diag}/dt\). The early post-bounce peak reaches \(1.07\times10^{53}\,\mathrm{erg\,s^{-1}}\) in the Lorentz-work term, \(8.06\times10^{52}\,\mathrm{erg\,s^{-1}}\) in \(\dot{Q}_{\rm diss}\), and \(7.04\times10^{52}\,\mathrm{erg\,s^{-1}}\) in \(dE_{\rm rot}/dt\). In time, the peak in magnetic energy occurs first, followed by the Lorentz-work term \((\mathbf{v}\cdot\mathbf{f}_L)\), and finally the dissipation term \(\dot{Q}_{\rm diss}\), indicating a progression from magnetic-field amplification to energy transfer to the flow and subsequent dissipation. The persistent positive residual indicates that a substantial fraction of the extracted magnetic energy is processed through effective dissipation associated with unresolved reconnection and sub-grid transfer, rather than being retained as resolved magnetic energy.}
    \label{fig:magnetic_power_budget}
\end{figure}

Fig.~\ref{fig:magnetic_power_budget} shows the corresponding power terms for B1R1. After the initial post-bounce peak, the resolved magnetic-energy derivative \(dE_B/dt\) remains comparatively small, while both the Lorentz-work term and the Poynting-flux term are substantial. A large positive residual, \(\dot{Q}_{\rm diss}\), is therefore required to close the budget. This indicates that magnetic energy extracted from differential rotation is not primarily retained as resolved magnetic energy, nor is it accounted for solely by the resolved electromagnetic energy flux through the control-volume boundary. Instead, a substantial fraction appears in the dissipation-like residual.

For comparison, we also overplot the time derivatives of the rotational energy and the diagnostic energy. At early times, the rotational reservoir is still affected by continued accretion and contraction of the PNS, and its evolution cannot be described by magnetic extraction alone. At later times, however, the picture becomes simpler: from \(\sim 0.3\) s after bounce onward, \(dE_{\rm rot}/dt\) becomes negative while the diagnostic energy continues to grow, and the magnitudes of \(dE_{\rm rot}/dt\) and \(dE_{\rm diag}/dt\) become comparable. This suggests that, once the early accretion-driven growth phase has subsided, continued magnetic extraction of rotational energy remains energetically coupled to the growth of the shock front. The arrows on the \(F_{\rm rot}\) curve in Fig.~\ref{fig:frot}, which indicate the Lorentz-work rate in the PNS, reinforce this interpretation by linking the decline of the rotational-energy reservoir directly to magnetic extraction.

The present diagnostics do not isolate the detailed pathway by which the dissipated magnetic energy contributes to the outflow. One plausible way is that part of this energy is thermalized in the inner flow and subsequently re-enters the outflow indirectly, for example through neutrino emission and heating. However, some of that energy may also be lost through escaping neutrinos or through thermal and hydrodynamic channels not separately tracked here. We therefore interpret the budget primarily as evidence that unresolved magnetic dissipation is dynamically important, while leaving the detailed partition of that energy to future work.

\subsubsection{Kink instability diagnostics}
\label{subsec:kink}

To test whether the magnetically driven polar outflow is destabilized by a non-axisymmetric MHD kink instability (the screw--pinch $m=1$ mode), we follow the magnetic-energy barycenter diagnostic introduced for global core-collapse simulations by \citet{mosta_magnetorotational_2014}. In a stack of fixed-$y$ planes we compute the energy-weighted barycenter of the magnetized spine,
\begin{equation}
\boldsymbol{\xi}(y,t) \equiv
\frac{\int \boldsymbol{x}_\perp\, B^2(\boldsymbol{x}_\perp,y,t)\, dA}
{\int B^2(\boldsymbol{x}_\perp,y,t)\, dA},
\qquad
r(y,t) \equiv |\boldsymbol{\xi}(y,t)|,
\label{eq:kink_barycenter}
\end{equation}
evaluated within a cylindrical aperture aligned with the rotation axis. A growing lateral displacement $r(y,t)$, together with coherent rotation of $\boldsymbol{\xi}$ in the transverse plane, is the canonical signature of an $m=1$ kink instability. We use this quantity because the lateral displacement of the magnetic-pressure barycenter provides a direct measure of the $m=1$ non-axisymmetric distortion of the magnetized outflow center: in the absence of a strong kink, the barycenter remains close to the rotation axis, while coherent growth of $r(y,t)$ indicates that the magnetized core of the outflow is being displaced away from that axis. 

\begin{figure}[t]
  \centering
  \includegraphics[width=\linewidth]{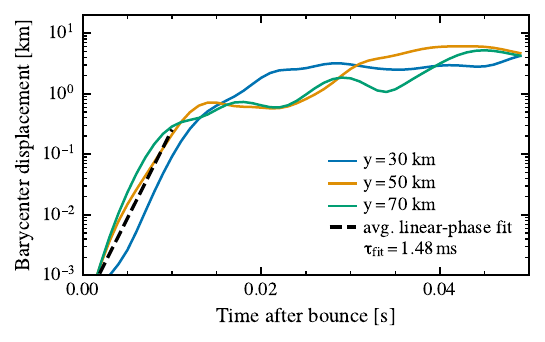}
  \caption{
  Kink-instability diagnostic for the magnetized rotating model B1R1. In fixed-$y$ planes we compute the magnetic-energy barycenter $\xi(y,t)$ and plot the lateral displacement $r(y,t)=|\xi(y,t)|$ within a cylindrical aperture aligned with the rotation axis. Shown are selected low-height probes together with an exponential fit to the early growth phase. The barycenter trajectories are lightly smoothed in time to suppress high-frequency fluctuations. The fitted e-folding time is $\tau_{\rm fit}$, obtained from the interval $t_{\rm pb}\simeq 0$--$10\,\mathrm{ms}$ and averaged over the lowest probes. This early exponential growth is qualitatively consistent with the onset of a kink non-axisymmetric instability.}
  \label{fig:kink_ryt_zoom}
\end{figure}

Fig.~\ref{fig:kink_ryt_zoom} highlights the early-time growth of the lateral barycenter displacement at selected low heights. Following previous work, we explicitly fit the initial growth phase with an exponential form, using the interval $t_{\rm pb}\simeq 0$--$10\,\mathrm{ms}$ and averaging over the lowest probes. This yields an empirical e-folding time $\tau_{\rm fit}$ of order a few milliseconds, indicating that the displacement grows rapidly before saturating.

As presented in \cite{mosta_magnetorotational_2014}, a useful comparison is provided by the expected growth time of the fastest-growing screw--pinch kink mode. For a toroidally dominated flow, this timescale is of order the Alfv\'en crossing time around the jet,
\begin{equation}
\tau_{\rm kink} \sim \frac{4 a \sqrt{\pi \rho}}{B_{\rm tor}},
\end{equation}
where $a$ is the characteristic radius of the magnetized core, $\rho$ is the local density, and $B_{\rm tor}$ is the toroidal magnetic-field strength. The mode is expected to become dynamically important once the unstable column satisfies the Kruskal--Shafranov condition,
\begin{equation}
\frac{B_{\rm tor}}{B_{\rm pol}}
\gtrsim
\frac{2\pi a}{L},
\label{eq:ks_criterion}
\end{equation}
where $L$ is the effective vertical length of the column and $B_{\rm pol}$ denotes the field component along it. Using representative conditions near the base of the outflow during the early post-bounce phase, we obtain a theoretical fast-growing-mode timescale, $\tau_{\rm fgm}$, of order a millisecond to a few milliseconds, broadly consistent with $\tau_{\rm fit}$.

We emphasize, however, that this estimate is only approximate in our global simulation. The flow is neither a stationary cylindrical pinch nor uniformly magnetized, and the inferred $\tau_{\rm kink}$ depends sensitively on the choice of the characteristic radius $a$ and the effective vertical length scale $L$ entering the instability criterion. Reasonable variations in these quantities can shift the theoretical estimate by factors of a few. We therefore do not interpret agreement between $\tau_{\rm fit}$ and $\tau_{\rm kink}$ too literally; rather, the key point is that the observed growth occurs on the timescale expected for a kink-like instability under plausible local conditions.

Within this diagnostic framework, our results can be directly compared to previous 3D magnetorotational studies. In \citet{mosta_magnetorotational_2014}, the early exponential growth of the barycenter displacement is likewise interpreted as evidence for a kink-like instability and coincides with the loss of a narrow outflow morphology. \citet{obergaulinger_magnetorotational_2021} similarly report non-axisymmetric distortions of the outflow spine, but find that the outflow can nevertheless propagate to large radii; in their magnetorotational models, rapid outflow propagation and toroidal-field shear limit fully disruptive kink instability growth. 

\begin{figure}[t]
  \centering
  \includegraphics[width=\linewidth]{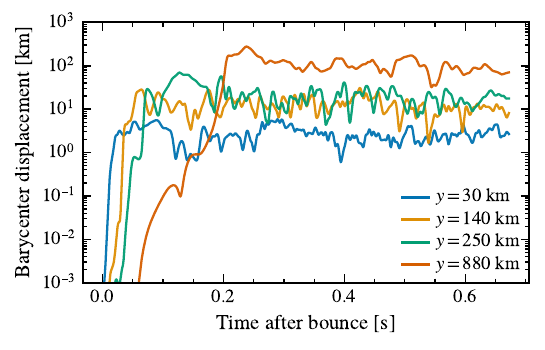}
  \caption{
  Lateral barycenter displacement $r(y,t)=|\boldsymbol{\xi}(y,t)|$ for many fixed-$y$ planes in the magnetized rotating model (B1R1), shown over the full time interval. The curves correspond to the sampled heights $y=\,$30, 140, 250, and 880 km, as indicated in the legend. The barycenter trajectories are lightly smoothed in time to suppress small-scale temporal fluctuations. Larger heights show a delayed onset and generally larger late-time amplitudes.}
  \label{fig:kink_ryt_full}
\end{figure}

Fig.~\ref{fig:kink_ryt_full} then shows the evolution of $r(y,t)$ across a broader range of heights over the full available time interval. Overall, the displacement becomes appreciable first at smaller $y$, while at larger $y$ the onset is delayed and the late-time amplitudes become larger. This systematic trend with height is consistent with a kinked, toroidally dominated funnel where a non-axisymmetric deformation develops near the base of the outflow and then strengthens as the outflow propagates outward, rather than maintaining a straight, narrow channel.

Our model shows the same early-time exponential growth of the barycenter displacement reported in previous kink diagnostics: we observe clear early-time exponential growth consistent with a kink-like mode, but without immediate catastrophic disruption of the outflow. The initially jet-like structure is, however, strongly quenched at later times (Fig.~\ref{fig:slices}), and the barycenter deviations quantified here likely contribute to that loss of a narrow, coherent outflow channel.
 
\subsubsection{Tilt and precession of the outflow and PNS}
\label{subsec:tilt}
\begin{figure}[t]
  \centering
  \includegraphics[width=\linewidth]{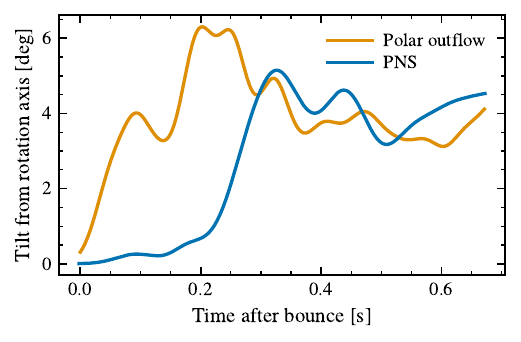}
  \caption{Tilt angle of the polar outflow axis and of the PNS angular-momentum axis relative to the original rotation axis in model B1R1. The outflow-axis direction is inferred from the barycenter/precession analysis in a plane located $70\,\mathrm{km}$ north of the center. The tilt curves are smoothed over a 20\,ms window to emphasize large-scale variations. The outflow develops an early, time-dependent misalignment of several degrees, while the PNS angular-momentum axis responds later and more weakly. The two tilt measures are therefore suggestive of a dynamical connection between the central object and the kink-distorted funnel, but they do not imply strict directional locking between the PNS and outflow axes.}
  \label{fig:tilt}
\end{figure}

The same non-axisymmetric dynamics that distort the polar funnel also lead to a measurable misalignment between the instantaneous outflow direction and the original rotation axis. To quantify this, we define an outflow-axis direction from the barycenter/precession analysis performed in a plane located $70\,\mathrm{km}$ north of the center and compare it to a PNS-axis proxy inferred from the direction defined by the PNS angular-momentum, $\vec{J}$. Fig.~\ref{fig:tilt} shows the corresponding tilt angles, measured relative to the original rotation axis, as a function of time. The outflow axis departs from the rotation axis early and reaches tilts of several degrees, while the PNS-axis proxy responds more gradually. The delayed evolution suggests that the PNS tilt is not driven directly by the same asymmetry that shapes the polar funnel. Rather, the coupling appears to be sequential: magnetic stresses modify the outflow geometry, the altered outflow then reorganizes the accretion flow, and the resulting changes in the angular momentum of the accreted material subsequently influence the PNS orientation. This is not unexpected: although the magnetic outflow is launched from the PNS region, its direction at larger radii is not simply determined by the bulk tilt of the PNS angular-momentum vector. The PNS angular-momentum evolution is shaped strongly by asymmetric accretion, whereas the measured outflow axis is additionally affected by magnetic stresses, precession, bending, and kink-like distortions as the funnel propagates outward. Moreover, because the accretion geometry itself is influenced by the outflow at larger heights, the coupling between PNS tilt and outflow tilt is inherently indirect and need not produce perfect alignment. Thus, while the tilt evolution indicates a dynamical connection between the central object and the non-axisymmetric outflow, it should not be interpreted as evidence for one-to-one tracking between the two axes. Nevertheless, toward the end of the simulation the outflow direction generally appears to precess around the more slowly varying PNS spin-axis, suggesting a loose late-time coupling between the two orientations rather than strict alignment.

\subsection{Ejecta composition and nucleosynthesis}
\label{sec:nuc}

We post-process models B1R1 and B0R0 to assess the composition of the unbound (as of the end of the respective simulations) mass. We omit B0R1 from this analysis due to the aforementioned issue (\S~\ref{subsec:resolution}) at late times with the neutrino transport that will impact the composition of our trajectories directly. For this we follow the methods presented in \cite{eggenberger_andersen_black_2025}, which are based on the backward trajectory tracing methods described in \cite{sieverding_tracer_2023}. In brief, we trace $\mathcal{O}(10^5)$ trajectories for each model starting at the end time for our simulations backwards in time until the trajectory either reaches $T=6$\,GK or the beginning of the simulation. We then forward process each resulting trajectory using SkyNet \citep{lippuner_skynet_2017} starting with either NSE (if the trajectory begins at $T=6$\,GK) or the original composition otherwise. We do not include neutrino interactions within SkyNet. We justify this by tracking the inferred $Y_e$ change from the neutrino interactions as determined by the in situ transport (but excluding any changes from the hydrodynamics) along the trajectory below $T=6$\,GK and find that $|\Delta Y_e|<0.001$ for more than 98\% of our trajectories and $|\Delta Y_e| \sim 0.0025$ at most. This would give very little impact on the evolution of $Y_e$ and the broad outcomes of the nucleosynthesis we are investigating here. For detailed calculations, especially of processes like the $\nu-p$ process, neutrinos are key. 

Given the uncertainty in the future trajectories of each tracer particle, we do not extrapolate the density and temperature of the trajectory forward in time but rather quote the compositions as of the end of the simulated time. However we note that further nucleosynthesis, especially of heavier elements in neutron rich trajectories, may occur. Additionally, given the large overburden of this progenitor, it may be that case that these currently unbound mass elements later become bound. \cite{eggenberger_andersen_black_2025} found that black hole supernovae, like the B0R0 model in our study, can rebind up to all of the synthesized elements in the 100s of seconds following black hole formation. Preliminary simulations of B0R0 past black hole formation confirm this fate for this model. However, the long term fates of B1R1 and B0R1, which have lower PNS masses as of the end of the simulations are unknown.

\begin{figure}[h!]
    \centering
    \includegraphics[width=\columnwidth]{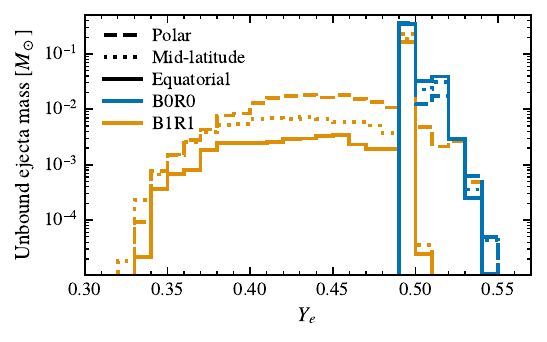}
    \caption{Distribution of unbound mass as a function of electron fraction for models B0R0 and B1R1 at the end of each simulation. We divide the unbound mass into three angular zones of equal solid angle: polar (dashed lines; $|\cos\theta| > 2/3$), equatorial (solid lines; $|\cos\theta| < 1/3$), and mid-latitude (dotted lines; intermediate between the two). Matter is classified as unbound when both the diagnostic energy and the radial velocity are positive. }
    \label{fig:yedistro}
\end{figure}

Before presenting the nucleosynthetic yields, we will first discuss the composition distribution of the ejecta. In Fig.~\ref{fig:yedistro}, we show the distribution of the unbound mass (as of the end of the simulation) as a function of $Y_e$ for B0R0 and B1R1, divided into three equal solid angle regions. The polar region (dashed lines; defined as $|\mathrm{cos}(\theta)|>2/3$), the equatorial region (solid lines; defined as $|\mathrm{ccos}(\theta)|<1/3$), and the remaining, mid-latitude region between the polar and the equatorial region. For each trajectory we take the starting value of $Y_e$ along the hydrodynamic trajectory as numerical mixing can smooth out the distribution \citep{stockinger_three-dimensional_2020}. There is a large peak at $Y_e\sim0.5$ that corresponds to the shocked ejecta that is swept up and unbound by the shock at large radii but never achieves high enough temperatures to undergo any nucleosynthesis.

\begin{figure}[h!]
    \centering
    \includegraphics[width=\columnwidth]{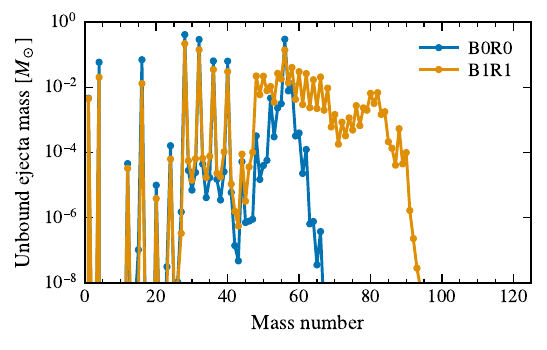}
    \caption{Distribution of unbound mass as a function of mass number for B0R0 and B1R1. The larger unbound mass for the lower atomic mass elements of the B0R0 model is due to the longer evolution time which leads to a larger unbound mass at this time.}
    \label{fig:massnumberdistro}
\end{figure}

The B0R0 simulation shows the expected distribution of ejecta for delayed neutrino driven supernovae, i.e. proton rich ejecta with $Y_e$ values between 0.5 and 0.55 \citep{wanajo_physical_2018,wang_nucleosynthetic_2024}. This is in contrast to very low compactness progenitor stars that can have a relatively fast shock expansion phase immediately following core bounce and show some neutron rich ejecta \citep{wanajo_nucleosynthesis_2018,stockinger_three-dimensional_2020,sandoval_three-dimensional_2021,wang_nucleosynthetic_2024}. Due to the lack of rotation, we expect and indeed find no systematic dependence of the $Y_e$ of the ejecta on the ejecta direction, although we can see even though the three regions have the same solid angle, there is a uneven distribution of the proton-rich ejecta that is consistent with the asymmetric shock expansion predominately occurring along the $+x$ direction (see Fig.~\ref{fig:slices}; bottom-right).

The B1R1 model has a sizable neutron-rich ejecta component, extending down to $Y_e \sim 0.33$, due to the fast development of the outflow, consistent with magnetorotational simulations in the literature \cite{mosta_r-process_2018, zha_nucleosynthesis_2024, reichert_magnetorotational_2023}. The relatively weak initial magnetic field, together with kink-induced broadening and destabilization of the polar outflow, prevents the production of ejecta with $Y_e < 0.32$, which are found in models with more extreme magnetic fields \cite{reichert_magnetorotational_2023}. Along the polar direction, especially inside the jetted material, $Y_e$ can reach values $>0.5$. This is due to the intense neutrino field along the polar direction. Along other directions, the $Y_e$ is generally restricted to less than 0.5 (except for the far-out, shocked, but generally unprocessed material that maintains $Y_e\sim0.5$).

In Fig.~\ref{fig:massnumberdistro}, we show the distribution (in mass number) of synthesized elements, as of the end of the simulations. Both simulations show a distribution of lower mass elements consistent with the material swept up by the expanding shock but otherwise unprocessed from their initial compositions. Explosive nucleosynthesis occurs at higher mass numbers. The B0R0 simulation, having only material with $Y_e>0.5$ gives synthesis up to $A\sim60$ with a peak at $A=56$, which corresponds to $\sim 0.3\,M_\odot$ of $^{56}$Ni at this early time. The B1R1 simulation, having more neutron rich ejecta synthesizes elements up to $A\sim90$ via a weak r-process,  but also has a peak at $A=56$, corresponding to $\sim 0.13\,M_\odot$ of $^{56}$Ni.

\section{Conclusions}
\label{sec:conclusions}

We have presented three-dimensional, (M)HD core-collapse calculations of a massive, extremely high-compactness ($39\,M_\odot$) progenitor \citep{aguilera-dena_precollapse_2020} to isolate how rotation and magnetic fields alter shock revival, outflow formation, and PNS evolution. By comparing a rotating magnetized model B1R1, a rotating non-magnetized model B0R1, and a non-rotating non-magnetized baseline B0R0, we find that all three models reach runaway shock expansion within the simulated interval, but through qualitatively different pathways. B1R1 develops the earliest sustained expansion and the clearest bipolar morphology; B0R0 follows with a broader plume-dominated neutrino-driven outflow and then collapses to a black hole by the end of the simulation; B0R1 revives latest and, within the simulated interval, remains at an earlier and more compact stage of shock development.

A central result is that magnetic fields in B1R1 do more than shape the outflow geometry. Maxwell stresses transport angular momentum out of the inner PNS, reduce centrifugal support, flatten the inner rotation profile, and tap the free energy of differential rotation. The angular-momentum budget shows that continued inward advection of angular momentum is strongly offset by magnetic torques, with neutrino angular-momentum losses providing a smaller secondary sink. At late times, the decline of the total rotational-energy reservoir occurs concurrently with continued growth of the diagnostic energy, consistent with magnetic extraction of energy from differential rotation contributing to the outflow. Both rotating models also show a modest late-time decline in enclosed PNS mass, reflecting a combination of redistribution across the adopted density boundary and genuine mass loss from the immediate PNS vicinity, most clearly in B1R1.

At the same time, the magnetic-energy budget shows that this extracted energy is not primarily retained as resolved magnetic energy, nor fully captured by the resolved Poynting flux through the control-volume boundary. Instead, a substantial fraction passes through the effective dissipation term $\dot{Q}_{\rm diss}$. We therefore regard unresolved magnetic dissipation as dynamically important in this model. These diagnostics do not isolate how that dissipated energy is partitioned between local heating, neutrino emission, and mechanical work on the outflow, so our complete understanding of the full energy pathway remains unresolved.

A second main result is that the magnetically dominated polar outflow in B1R1 is not a steady, narrowly collimated jet. The barycenter-displacement diagnostic shows early exponential growth on the few-millisecond timescale expected for a kink instability, and the funnel subsequently becomes broader, time dependent, and tilted by several degrees relative to the original rotation axis. The outflow survives and propagates to large radii, but its morphology is better described as a kink-distorted magnetically dominated polar outflow than as a persistent narrow jet. This is precisely where fully three-dimensional modelling matters most: axisymmetry can overestimate both the stability and the collimation of the outflow. 

The ejecta analysis suggests that magnetic support alters composition as well as dynamics. The same non-axisymmetric shock geometry also feeds back on the central object by redirecting accretion and tilting the PNS angular-momentum axis by $\sim 5^\circ$ away from the original rotation axis. B0R0 shows the proton-rich ejecta expected for a delayed neutrino-driven outflow, whereas B1R1 contains a substantial neutron-rich component extending down to $Y_e\sim0.33$. In the currently unbound material, this corresponds to nucleosynthesis reaching $A\sim90$, while B0R0 remains concentrated near iron-group production. These yields should be treated as provisional, however, because we do not extrapolate the thermodynamic trajectories beyond the simulated interval and the long-term bound or unbound fate of the ejecta is not yet known. In addition, the large binding energy of the remaining stellar overburden means that some material may fall back at later times, potentially leading to delayed black-hole formation. This further underscores the need for longer-term simulations to determine the final remnant fate and asymptotic ejecta properties.

Several limitations define the next steps. First, the achievable resolution remains a central limitation of this study. Global 3D MHD simulations with multidimensional neutrino transport are computationally expensive, which restricts the spatial resolution, simulated duration, and size of the model suite. This limitation affects both the physical interpretation of B1R1 as an exploratory strong-field scenario, where the imposed large-scale seed field stands in for amplification processes that cannot yet be followed self-consistently, and the degree to which small-scale MHD instabilities can be resolved.

Second, several quantitative diagnostics, including the inferred effective dissipation and PNS moment of inertia, are resolution sensitive and require controlled convergence studies. Code comparisons targeted specifically at magnetic-energy budgets, effective dissipation, and angular-momentum transport would be particularly valuable for assessing which aspects of these diagnostics are physical and which depend on numerical method and resolution.

Third, the late-time evolution of B0R1 is affected by a transport inconsistency and should not be over-interpreted. Fourth, the ultimate fates of B1R1 and B0R1 require substantially longer simulations.

Finally, although parts of the flow satisfy local MRI-resolvability estimates, the present resolution is not sufficient to establish MRI-driven turbulence or MRI-mediated field amplification as the primary driver of the evolution. The diagnostics presented here can be explained by magnetic winding, Maxwell stresses, Lorentz work, and Poynting-flux transport, while a definitive assessment of the MRI would require higher-resolution simulations and dedicated convergence tests.

Overall, these simulations support a picture in which rapid rotation together with strong magnetic fields can open an early magnetically aided pathway to shock revival in an extremely compact progenitor, but one whose evolution and outcome are inherently three-dimensional. The same magnetic stresses that help launch the polar outflow also spin down the inner core, broaden and distort the funnel, and route a substantial fraction of the extracted energy through unresolved dissipation. Any connection between magnetorotational core collapse and engine-driven transients therefore depends not only on whether a polar outflow is launched, but also on how that outflow survives, bends, and reorganizes in full 3D.
\FloatBarrier

\begin{acknowledgements}
We thank Andrei Beloborodov, Dhrubaditya Mitra, Axel Brandenburg, Matteo Bugli, J\'er\^ome Guilet, and Philipp M\"osta for valuable discussions and insights that helped shape this work. We are especially grateful to the broader core-collapse supernova and compact-object community for ongoing exchanges that informed many aspects of this study. This work is supported by the Swedish Research Council (Project No. 2020-00452). The computations were enabled by resources provided by the National Academic Infrastructure for Supercomputing in Sweden (NAISS), partially funded by the Swedish Research Council through grant agreement no. 2022-06725.

This work made use of the open-source analysis and visualization packages \texttt{yt} \citep{turk_yt_2011} and \texttt{Matplotlib} \citep{hunter_matplotlib_2007}.

\end{acknowledgements}

\bibliographystyle{aa}
\bibliography{references}

\appendix

\section{Derivation of the Angular Momentum Flux (Torque) Equation}
\label{app:torque}

In this appendix we derive the angular-momentum balance equation for a PNS from the ideal-MHD momentum equation and then add the neutrino contribution. Throughout, we use cgs units and adopt a fixed spherical control volume $V$ of radius $R$, bounded by the surface $\partial V$. 

Let's start with the momentum equation:
\begin{equation}
\frac{\partial(\rho \mathbf v)}{\partial t} + \nabla\cdot(\rho \mathbf v \mathbf v) = -\nabla p - \rho\nabla\Phi + \frac{1}{4\pi}(\nabla \times \mathbf B)\times \mathbf B
\end{equation}
where $\Phi$ is the (central) gravitational potential.

For angular momentum in a fixed volume:
\begin{equation}
\mathbf J = \int_V \mathbf r \times \rho\mathbf v\, d V
\end{equation}

Taking the time derivative and substituting the momentum equation gives:

\begin{equation}
\frac{\partial \mathbf J}{\partial t} = \int_V \mathbf r \times \left[- \nabla\cdot(\rho \mathbf v \mathbf v) + \frac{1}{4\pi}(\nabla \times \mathbf B)\times \mathbf B \right]\,d V
\end{equation}

The pressure and gravitational terms do not contribute to the net torque about the origin for a spherical control volume centered on the origin: $\Delta P$ ..., and $\Delta \Phi$ is radial for a central potential. We therefore retain only the advective and magnetic contributions.

Using the vector identity
\begin{equation}
(\nabla \times \mathbf B)\times \mathbf B = \nabla \cdot (\mathbf B \mathbf B) - \nabla (B^2/2).
\end{equation}
The gradient term again produces no net torque about the origin, so the expression becomes

\begin{equation}
\frac{\partial \mathbf J}{\partial t} = \int_V \mathbf r \times \left[ - \nabla\cdot(\rho \mathbf v \mathbf v) + \frac{1}{4\pi}\nabla \cdot (\mathbf B \mathbf B) \right]\,d V.
\end{equation}

Next we use the identity
\begin{equation}
\mathbf r \times [\nabla \cdot \mathbf A] = \nabla \cdot [\mathbf r \times \mathbf A],
\end{equation}

where $\mathbf A$ is a rank-2 tensor and the divergence acts on $\mathbf A$.

Applying this identity to the advective and magnetic flux tensors gives

\begin{equation}
\frac{\partial \mathbf J}{\partial t} = \int_V \nabla \cdot \left( - \mathbf r \times (\rho \mathbf v \mathbf v) + \frac{1}{4\pi} \mathbf r \times (\mathbf B \mathbf B) \right)\, d V
\end{equation}

Using the divergence theorem, we convert the volume integral to a surface integral over $\partial V$:
\begin{equation}
\frac{\partial \mathbf J}{\partial t}
=
\oint_{\partial V}
\left[
-\mathbf r \times (\rho \mathbf v \mathbf v)
+\frac{1}{4\pi}\mathbf r \times (\mathbf B \mathbf B)
\right]\cdot d\mathbf A.
\label{eq:Jdot_surface_tensor_appendix}
\end{equation}

For a spherical surface of radius $R$, the outward area element is $d\mathbf A = \hat{\mathbf n}\,dA$ with $\hat{\mathbf n} = \mathbf r/R$.

Defining the radial components:
\[
v_{r} = \mathbf v \cdot \hat{\mathbf n}, \qquad
B_{r} = \mathbf B \cdot \hat{\mathbf n}
\]
Equation above becomes:
\begin{equation}
\boxed{
\frac{\partial\mathbf J}{\partial t}
=
-\oint_{\partial V}
\left[
(\mathbf r \times \mathbf v)\,\rho\,v_r
-\frac{1}{4\pi}(\mathbf r \times \mathbf B)\,B_r
\right] dA
}
\label{eq:Jdot_final_appendix}
\end{equation}

This result expresses the total torque (rate of change of angular momentum) through a spherical surface due to the hydrodynamic and magnetic angular momentum fluxes through that surface. 

Neutrinos will also feedback on the momentum of the fluid. Due to the rotation of the PNS the neutrinos leaving the surface will have have net angular momentum. Following \cite{epstein_neutrino_1978,harada_neutrino_2019}, the reduction of angular momentum due to the flux through the surface of sphere for a single neutrino species and energy group is given as,

\begin{equation}
\frac{\partial \mathbf{J}^{(\nu)}}{\partial t}
=
-\oint_{\partial V}
\left[
\mathbf{r} \times \bigl(\mathbf{P}^{(\nu)}\cdot \hat{\mathbf r}\bigr)
\right] dA,
\label{eq:djnudt_sphere}
\end{equation}
\noindent
where $\mathbf{P}^{(\nu)}$ is the second moment of the neutrino distribution function. Using the M1 closure form of the second moment along with the Minerbo closure relation we express
\begin{equation}
P^{(\nu)}_{ij} = \frac{3(1-\chi^{(\nu)})}{2}\frac{E^{(\nu)}\delta_{ij}}{3} + \frac{3\chi^{(\nu)}-1}{2}\frac{E^{(\nu)}F^{(\nu)}_iF^{(\nu)}_j}{|\mathbf{F}^{(\nu)}|^2}
\end{equation}
\noindent
where $\chi^{(\nu)}=1/3 + 2/15(3{f^{(\nu)}}^2-{f^{(\nu)}}^3+3{f^{(\nu)}}^4)$, with $f^{(\nu)}=|\mathbf{F}^{(\nu)}|/(cE^{(\nu)})$, $E^{(\nu)}$ is the neutrino energy density (erg/cm$^3$), and $F_i^{(\nu)}$ is the neutrino momentum density (erg/cm$^2$/s). The optically thick term represents an isotropic pressure and does not contribute to the angular momentum flux. The optically thin term does. With this closure, Eq.~\ref{eq:djnudt_sphere} becomes,
\begin{equation}
\frac{\partial J_i^{(\nu)}}{\partial t}
=
-\oint_{\partial V}
\left[
\epsilon_{ijk}\, r_j\,
\frac{3\chi^{(\nu)}-1}{2}
\frac{E^{(\nu)} F^{(\nu)}_k F^{(\nu)}_l \hat{r}_l}{|\mathbf{F}^{(\nu)}|^2}
\right] dA.
\label{eq:djnudt}
\end{equation}
\noindent
Restoring vector notation and explicitly summing over the different neutrino species and energy groups,
\begin{equation}
\frac{\partial \mathbf{J}^{(\nu)}}{\partial t} = -\sum_{i:\{\nu_e,\bar{\nu}_e,\nu_x\}} \sum_{j=1}^{12}
        \int_{A}\Bigl[
            (\mathbf{r} \times \mathbf{F}^{(\nu_{ij})}) (\mathbf{\hat{r}}\cdot \mathbf{F}^{(\nu_{ij})})\frac{3\chi^{(\nu_{ij})}-1}{2}\frac{E^{(\nu_{ij})}}{|\mathbf{F}^{(\nu_{ij})}|^2}
        \Bigr]\ dA \,.
        \label{eq:djnudt_sum}
\end{equation} 

\section{Resolution sensitivity of the PNS moment of inertia}
\label{app:moi_resolution}

The PNS moment of inertia is computed from the mass distribution within the high-density core (here using $\rho \ge 10^{11}\,\mathrm{g\,cm^{-3}}$) and therefore depends on how well the simulation resolves the core structure and its gradients. In practice, we find that the inferred moment of inertia is noticeably sensitive to the grid resolution.

\begin{figure}[!htbp]
  \centering
  \includegraphics[width=\columnwidth]{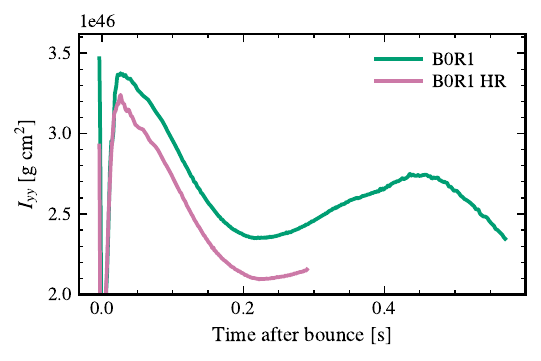}
  \caption{
  Resolution comparison of the PNS moment of inertia about the rotation axis, $I_{yy}$, between the baseline B0R1 model and its higher-resolution (HR) counterpart. While the qualitative time evolution is similar, the normalization and finer features are resolution sensitive. At the final common time, the HR value is lower than the baseline value by $\sim 12$\%.
  }
  \label{fig:app_moi_res}
\end{figure}

Fig.~\ref{fig:app_moi_res} compares the baseline realization of the B0R1 model with its higher-resolution B0R1 (HR) counterpart. Although the two simulations show broadly similar post-bounce trends, the overall magnitude and short-timescale variations of $I_{yy}$ (the moment of inertia about the rotation axis) differ between the baseline and HR models. At the final common time, the HR model has a moment of inertia that is lower by $\sim 12$\% relative to the baseline model. This sensitivity is consistent with the fact that $I_{yy}$ weights material by the squared distance to the rotation axis and is thus especially responsive to modest changes in the density profile and effective PNS radius. 

\begin{figure}[!htbp]
  \centering
  \includegraphics[width=\columnwidth]{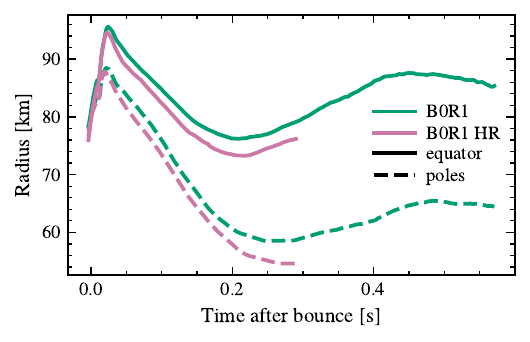}
  \caption{
  Resolution comparison of the equatorial and polar-averaged PNS radii for the baseline B0R1 model and its higher-resolution (HR) counterpart. The two simulations show similar qualitative evolution, but the HR model is systematically more compact. At the final common time, the HR equatorial radius is smaller by $\sim 3$\%, while the HR polar-averaged radius is smaller by $\sim 7$\%.
  }
  \label{fig:app_radius_res}
\end{figure}

We also compare the polar and equatorial PNS radii inferred from the same density threshold (Fig.~\ref{fig:app_radius_res}). The baseline and HR simulations show broadly consistent trends in both $R_{\rm pol}$ and $R_{\rm eq}$, but with non-negligible offsets that track the differences in $I_{yy}$. At the final common time, the HR model is more compact by $\sim 3$\% in equatorial radius and by $\sim 7$\% in the polar-averaged radius. The larger fractional change in $I_{yy}$ reflects the additional lever-arm weighting in the moment-of-inertia integral.

Together, these comparisons indicate that quantitative statements based on the absolute value of the moment of inertia (and closely related rotational diagnostics) should be interpreted with some caution at the present resolution.

A systematic resolution study is beyond the scope of this work. Nevertheless, the baseline--HR differences in Figs.~\ref{fig:app_moi_res} and~\ref{fig:app_radius_res} motivate higher-resolution simulations as a clear direction for future work, both to improve convergence of the PNS structural diagnostics and to reduce uncertainty in derived rotational measures that depend on $I$.

\section{Definition of $Q_{diss}$}
\label{app:diss_deriv}

We define $Q_{\mathrm{diss}}$ from the magnetic-energy equation. Starting from the induction equation
\begin{equation}
\frac{\partial \mathbf{B}}{\partial t} = - c\, \nabla \times \mathbf{E},
\end{equation}

and the magnetic-energy density
\begin{equation}
e_B = \frac{B^2}{8\pi}.
\end{equation}

we obtain
\begin{equation}
\frac{\partial e_B}{\partial t}
=
\frac{1}{4\pi}\mathbf{B}\cdot\frac{\partial \mathbf{B}}{\partial t}
=
-\frac{c}{4\pi}\mathbf{B}\cdot(\nabla\times\mathbf{E}).
\end{equation}

Using the vector identity for the divergence of a cross product,
$\nabla \cdot (\mathbf{A} \times \mathbf{B}) = \mathbf{B} \cdot (\nabla \times \mathbf{A}) - \mathbf{A} \cdot (\nabla \times \mathbf{B})$,

together with Amp\`ere's law,
\begin{equation}
\nabla\times\mathbf{B} = \frac{4\pi}{c}\,\mathbf{J},
\end{equation}

gives
\begin{equation}
\frac{\partial e_B}{\partial t}
=
-\nabla\cdot\left(\frac{c}{4\pi}\,\mathbf{E}\times\mathbf{B}\right)
-\mathbf{J}\cdot\mathbf{E}.
\end{equation}

We then define the Poynting flux
\begin{equation}
\mathbf{S} \equiv \frac{c}{4\pi}\,\mathbf{E}\times\mathbf{B}.
\end{equation}

the magnetic-energy equation becomes
\begin{equation}
\frac{\partial e_B}{\partial t} + \nabla\cdot\mathbf{S} = -\mathbf{J}\cdot\mathbf{E}.
\label{eq:mag_energy_app}
\end{equation}

To separate reversible magnetic work from irreversible loss, we use the resistive-MHD Ohm's law
\begin{equation}
\mathbf{E} + \frac{\mathbf{v}\times\mathbf{B}}{c} = \eta \mathbf{J},
\end{equation}
so that
\begin{equation}
\mathbf{J}\cdot\mathbf{E}
=
-\frac{1}{c}\,\mathbf{J}\cdot(\mathbf{v}\times\mathbf{B}) + \eta J^2.
\end{equation}

Using the scalar triple product identity
\begin{equation}
-\frac{1}{c}\,\mathbf{J}\cdot(\mathbf{v}\times\mathbf{B})
=
\mathbf{v}\cdot\left(\frac{\mathbf{J}\times\mathbf{B}}{c}\right),
\end{equation}
and defining the Lorentz-force density
\begin{equation}
\mathbf{f}_L \equiv \frac{\mathbf{J}\times\mathbf{B}}{c},
\end{equation}
we find
\begin{equation}
\mathbf{J}\cdot\mathbf{E}
=
\mathbf{v}\cdot\mathbf{f}_L + \eta J^2.
\end{equation}

Substituting this into Eq.~\ref{eq:mag_energy_app} yields
\begin{equation}
\frac{\partial e_B}{\partial t} + \nabla\cdot\mathbf{S}
=
-\mathbf{v}\cdot\mathbf{f}_L - \eta J^2.
\end{equation}
We therefore define
\begin{equation}
Q_{\mathrm{diss}}(\mathbf{x},t)
\equiv
-\left(
\frac{\partial e_B}{\partial t}
+
\nabla\cdot\mathbf{S}
+
\mathbf{v}\cdot\mathbf{f}_L
\right),
\label{eq:qdiss_def}
\end{equation}
so that, in resistive MHD,
\begin{equation}
Q_{\mathrm{diss}} = \eta J^2.
\end{equation}

Equation~\ref{eq:qdiss_def} shows that $Q_{\mathrm{diss}}$ is the part of the magnetic-energy loss that cannot be accounted for by Poynting-flux transport or by reversible mechanical work by the Lorentz force. In a system with explicit resistivity, it reduces to the usual Ohmic heating rate, \(\eta J^2\), where \(\eta\) is the Ohmic resistivity in Gaussian cgs units. In ideal MHD, the analytic value is $Q_{\mathrm{diss}}=0$, so any non-zero value measured from the discrete simulation should be interpreted as an effective dissipation associated with numerical resistivity and unresolved small-scale magnetic structure.

\end{document}